\documentclass[]{aastex631}
\usepackage{afterpage,booktabs}

\newcommand\aastex{AAS\TeX}
\newcommand\latex{La\TeX}

\usepackage{xcolor}
\usepackage{soul}
\usepackage{appendix}
\newcounter{tr}
\ifnum \value{tr}>5

\begin{document}
\makeatletter
\def\@submitted{Accepted by ApJ: July 06 2025}
\makeatother

\title{Gamma-ray burst pulse structures and emission mechanisms}

\author[0009-0006-7782-0164]{A Gowri}
\affiliation{Indian Institute of Science Education and Research (IISER), Mohali, India}
\affiliation{Department of Physics, Bar-Ilan University, Ramat-Gan 52900, Israel}

\author[0000-0001-8667-0889]{A. Pe{'}er}
\affiliation{Department of Physics, Bar-Ilan University, Ramat-Gan 52900, Israel}

\author[0000-0001-8667-0889]{F. Ryde}
\affiliation{Department of Physics, KTH Royal Institute of Technology and The Oskar Klein Centre \\ 
SE-10691 Stockholm, Sweden}

\author[0000-0002-8852-7530]{H. Dereli-B\'egu\'e}
\affiliation{Department of Physics, Bar-Ilan University, Ramat-Gan 52900, Israel}





\begin{abstract}


The prompt phase X- and $\gamma$-ray light curves of gamma-ray bursts (GRBs) exhibit erratic and complex behaviour, often with multiple pulses. The temporal shape of individual pulses is often modelled as `fast rise exponential decay' (FRED). 
Here, we introduce a novel fitting function to quantify pulse asymmetry. We conduct a light curve and a time-resolved spectral analysis on 61 pulses from 22 GRBs detected by the Fermi Gamma-ray Burst Monitor. 
Contrary to previous claims, we find that only $\sim 50$\% of pulse lightcurves in our sample show a FRED shape, while about 25\% have a symmetric lightcurve, and the other 25\% have a mixed shape. Furthermore,
our analysis reveals a clear trend: in multi-pulse bursts, the initial pulse tends to exhibit the most symmetric light curve, while subsequent pulses become increasingly asymmetric, adopting a more FRED-like shape. 
Additionally, we correlate the temporal and spectral shapes of the pulses.
By fitting the spectra with the classical ``Band" function, we find a moderate positive Spearman correlation index of 0.23 between pulse asymmetry and the low-energy spectral index $\alpha_{\max}$ (the maximum value across all time bins covering an individual pulse).
Thus, during GRB light curves, the pulses tend to get more asymmetric and spectrally softer with time. We interpret this as a transition in the dominant emission mechanism from photospheric (symmetric-like and hard) to non-thermal emission above the photosphere and show that this interpretation aligns with a GRB jet Lorentz factor of the order of a few 10s in many cases.
\end{abstract}


\section{Introduction} \label{sec:intro}

Gamma-ray bursts (GRBs) are well known to have erratic light curves in the X- and $\gamma$-ray bands during their prompt emission phase.
While the light curves of some GRBs are smooth (i.e., contain a single pulse), the light curves of many others contain multiple, sometimes overlapping pulses. 
These pulses are considered to be individual emission episodes varying largely in duration and energy. 
This makes it very hard to deduce the physical properties of GRBs, as a complete model requires understanding the origin of individual pulse duration, shape, spectrum, and the relation between the different pulses. 

Studies of individual pulses have proven fruitful in revealing some key properties. In early works using data from the BATSE instrument on the Compton Gamma-Ray Observatory, \citet{Norris1996, norris2005} and \citet{hakkila2018} found that individual pulses are typically characterized by a non-symmetric shape, which is modelled as `fast rise exponential decay' (FRED). 
This shape was found independent of the observed energy band \citep{hakkila2009}. A FRED pulse shape can result from a sudden dissipation of energy. This idea indeed suggests a non-symmetric pulse shape: 
While the rise in the light curve is due to the unspecified energy dissipation mechanism, the decay can be affected by additional physical effects, such as light aberration \citep[known as high-latitude emission; e.g.,][]{panaitescu2001}. 

Other efforts to quantify the pulse shapes include models with a power law rise and decay
\citep[e.g.,][]{Norris1996, lee2000}, with a power-law rise and exponential decay \citep[e.g.,][]{jia2005}, with a Gaussian rise and an exponential decay \citep[e.g.,][]{stern1996}, with a Gaussian rise and decay \citep{bhat2011}, and involving more complex rise and decay functions \citep[e.g.,][]{kocevski2003}.

Additional observational findings that may shed light on the physical origin of the pulses include:
(i) Pulses detected at lower energies tend to have longer duration compared to pulses observed at higher energies \citep{richardson1996}. (ii) A correlation was
found between pulse fluence and duration \citep{hakkila2011}: brighter pulses tend to be longer. And 
(iii) Pulse durations were found to be anti-correlated with their peak fluxes \citep{hakkila2011}. 

Along with these temporal analyses of the GRB pulses' light curves, \cite{li2021} employed a time-resolved spectral analysis method in analyzing multi-pulsed GRBs observed by \textit{Fermi}-GBM. They found a clear trend between the low energy spectral index $\alpha$  and the pulse number (first, second, third, etc.). As a burst evolves and the pulse number increases, the value of $\alpha$ decreases. This indicates a temporal evolution in the radiative mechanism across pulses during the prompt phase in a single burst. 

Most of the above-mentioned works dealt with analyzing data obtained from the BATSE instrument. However, some of the well-used models provide poor fits to data \citep{hintze2022} obtained by the much newer Gamma-ray Bursts Monitor (GBM) on board the Fermi satellite \citep{meegan2009}. Indeed, it seems that very few studies have been published analyzing the shape of pulses observed by the GBM instrument. Clearly, being a newer instrument, the signal-to-noise (S/N) ratio of the data provided by the GBM is higher than that of the BATSE. This motivates a study of the GRB pulse shapes detected by the GBM, aiming to propose a more physically grounded model for fitting these pulse shapes and to explore various potential correlations between pulse shape, pulse count, duration, and spectral characteristics.
Focusing on multi-pulse evolution enables one to study whether the changes in pulse structures correspond to changes in the emission mechanism within the relativistic flow.


According to the ``fireball" model of GRBs, energy dissipation processes in the highly relativistic jet that is ejected during the formation of a black hole produce the prompt emission \citep{rees1992, meszaros2002}. There are two leading emission mechanisms proposed in explaining the observed prompt emission signal, assuming it is of leptonic origin. 
The first is radiation from the photosphere, where the expanding jet becomes transparent \citep{goodman1986, rees2005, peer2006}. 
Alternatively, a region further from the progenitor, where the jet's kinetic or magnetic energy dissipation occurs, could be the location of observed photons. This dissipated energy is used in accelerating electrons to high energies, resulting in the emission of synchrotron radiation \citep{piran1993, sari1998, lloyd2000}, which may be followed by inverse Compton (IC) scattering at high energies. The observed timescales of these emission episodes are $t_{\gamma} = r_{\gamma}/\Gamma^2 c$, where $c$ is the speed of light, $\Gamma$ the Lorentz factor of the jet and $r_{\gamma}$ the radius of the emission site.   

By definition, the first photons are always observed from the photosphere, whose radius is expected in the range $r_{ph} \sim 10^{11} - 10^{13}$~cm, depending on the outflow parameters \citep[e.g.,][]{peer2012}. The expected spectrum is a modified black-body. A ``pure" black-body (Planck) function is not expected due to both light aberration \citep{peer2008} and possible energy dissipation below and close to the photosphere \citep{peer2006, giannios2006}. Energy dissipation at larger radii, $r_\gamma \approx 10^{13}-10^{17} $~cm, will occur in the optically thin region and be observed after a few to several hundred seconds. Being in the optically thin regime, the resulting spectra reflect the radiative process and are expected to be a combination of synchrotron at lower energies and IC at higher energies. In principle, in extended jetted outflows as is expected in long GRBs, the two emission models can coexist and contribute simultaneously to the observed spectra \citep[e.g.,][]{ryde2005, ryde2009, guiriec2011, ajello2020}.

The value of the low-energy spectral index ($\alpha$), which is typically well measured, can be used as a proxy for the emission mechanism: if the emission is of thermal (photospheric) origin, $\alpha$ represents the Rayleigh-Jeans part of the spectrum (although modified) and is therefore expected to be harder than the expected index from synchrotron emission. We point out that various effects, both physical, geometrical, and instrumental, will act to modify the observed value of $\alpha$, and therefore, the discrimination between the radiative models is done using empirical models fitted to the synthetic data. The boundaries of synchrotron emission would lie below $\alpha \le -0.66 $ (known as the ``synchrotron line-of-death"). 
On the other hand, it was shown by \cite{Acuner20} that $\alpha \ge -0.1$ is a clear indication of a non-dissipative photospheric (NDP) origin. This value is modified from the theoretical limit of $\alpha = +0.4$ \citep{beloborodov2011} constrained by the finite detector's energy band, the spectral fitting procedure (they used a cut-off power law), and the finite S/N ratio. In order not to be sensitive to these unknowns, we interpret $\alpha$ of indicating thermal origin using only the non-modified theoretical limit, $\alpha \ge +0.4$; see further discussion below.    


The spectral differences may very well be accompanied by temporal differences. If the emission radius is above the photosphere, the physical mechanisms that govern the rise time and the decay time of pulses will, in general, be different. For example, the rise time reflects the time in which energy (kinetic or magnetic) is dissipated, while the decay will result from different effects, such as the jet structure and geometry, which varies in time, the density, which varies with radius etc. On the other hand, emission from the photosphere is expected to be more symmetric, as (i) the photospheric radius is not expected to vary considerably during the pulse duration, and (ii) photons that are emitted below the photosphere diffuse through the plasma until they escape, causing a more symmetric pulse structure. Thus, overall, one can expect a correlation between the pulse shape and spectra. 



This paper aims at examining the pulse shapes in the GBM era. We introduce a new pulse fitting function, which is more flexible than the fit function used by \cite{Norris1996}. This enables us to capture a larger range of pulse shapes than before. We then look for various correlations that are found in the data and interpret them. 
As we show here, some clear correlations and trends exist, which may provide new insight into the physics of GRB jets.

The paper is organized as follows. Section 2 outlines the data analysis process, including sample selection, pulse modelling, and spectral analysis. The results following the analysis of the observational properties
are provided in Section 3. Finally, Section 4 includes a
discussion of the findings and the summary of our investigation. Throughout the paper, the standard $\Lambda$CDM
cosmology with the parameters $H_0 = 67.4 ~{\rm km~ s^{-1}~ Mpc^{-1}}~
,
\Omega_{M} = 0.315$, and $\Omega_{\Lambda} = 0.685$ are adopted \citet{planck2021}.

\section{Data Analysis} \label{sec:style}

\subsection{Sample selection}

We use data obtained by the {\it Fermi} Gamma-Ray Burst Monitor (GBM), which includes 12 NaI and 2 BGO detectors covering roughly the energy range between 8 keV to 40 MeV \citep{meegan2009}. 
We define the basic unit of a burst morphology as a `pulse', which is considered as an individual emission episode. 
This is part of the light curve that rises, reaches a peak, and decays.
%
%
Due to the variability in the pulse structures in GRBs, the method for pulse demarcations has been mostly visual in the previous studies. As the reproducibility of pulses is of prominent importance, we have made specific criteria for the count rate to ensure that the general pulse envelope can be identified.

\begin{itemize}
    \item Ideally, a pulse is defined as a distinct emission episode characterized by a rise and fall in the count-rate relative to the background level. The pulse start time is defined as the time at which the pulse count-rate sustainably rises above the background level. 
    The effects of Poissonian noise in pulse selection are taken care of by the application of a polynomial fitting to the background level and the Bayesian block binning method. This enables us to identify the beginning of the pulse. The pulse count rate rises until it reaches a peak. Following, the pulse count rate descends, marking the beginning of the decay phase. The pulse end time is set once the count rate reaches again the background level.
    \item  In some cases, the pulse count-rate does not start or end at background levels. This happens when one pulse immediately follows another pulse. In order to identify whether an emission episode forms an individual, independent pulse or should be considered as a fluctuation within an already identified pulse, the following criterion is applied. 
    We measure the count rates at the beginning and end of the emission episode.
     \begin{itemize}
        \item To consider the emission episode that immediately follows a pulse as another, independent pulse, we calculate the count rates at the start time of the episode and at its peak. Our criterion is that the count rate at the start time must be less than 50\%  of the peak count rate. If this criterion is met, we consider the episode as an independent pulse. Otherwise, we consider it as a continuation of the previous pulse or as a subdominant variation.
        \item  Similarly, the emission episode is considered a separated pulse if at the end time of the episode, the count rate is at most 50\% of the peak count-rate.     
    \end{itemize}     
\end{itemize}

To mitigate the effects of light-curve binning and background noisy flares in the initial data, a 2\% error is allowed in this criterion, see Section \ref{sec:pulsemodel}.
We denote the duration of a pulse as the source interval.

Furthermore, these criteria effectively act as a low-pass filter, enabling smooth fluctuations in the light curve. As suggested by \cite{li2021}, we are allowing pulses with smaller spikes whose heights are bounded by an approximate pulse-shaped envelope, identified after the 0.1s binning of the brightest NaI detector light curve. 
This avoids statistical variations that could be produced in the light curve due to background and ensures a high quality of individual pulses in a burst, for this analysis.

\begin{figure}
    \centering
    \includegraphics[width=12 cm]{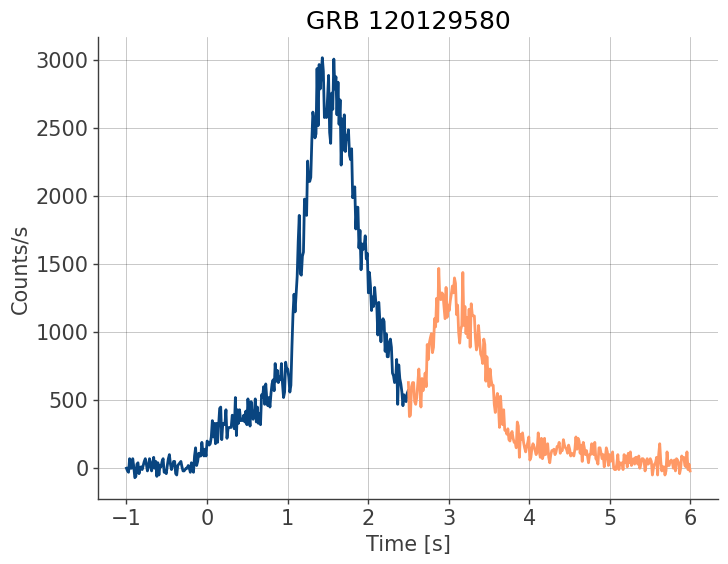}
    \caption{GBM light curve of GRB 120129 shows two pulses. The first pulse spans from -0.7~s to 2.5~s, followed by the second pulse from 2.5~s to 4~s. 
 }
    \label{fig:pulseselection}
\end{figure}

The different individual pulses within a burst may or may not be separated by quiescent intervals. In our sample, we choose only long GRBs with $T_{90} > 2 s$, that feature at least two different pulses in order to study the evolution of the pulse structure during the burst. 
An example of the sample selected for the study is given in Figure \ref{fig:pulseselection}.

After identifying the pulses, we impose additional selection criteria as follows. The light curves are binned using Bayesian blocks \citep{Scargle2013}; see further discussion in section \ref{sec:2.4} below. A minimum selection criterion is that each pulse in each GRB light curve has at least two time bins, each of them having at least a statistical significance $S\ge20$. An exact definition of $S$, which is a measure of the signal-to-noise ratio, can be found in \cite{Vianello2018b}. Secondly, we selected only pulses which have the coefficient of determination $r^2$ value $\ge$ 0.7 \footnote{
The coefficient of determination, denoted by $r^2$ $(0 \leq r^2 \leq 1)$ is a measure of how well a function fits a given data. It is defined by $r^2 \equiv 1 - {RSS}/{TSS}$,
where $RSS$ is the sum of squares of the fit residuals, and $TSS$ is the total sum of squares. In our analysis, the pulses are selected only if $r^2 > 0.7$. } after fitting with our novel pulse shape function. 
We also removed the fits with significant errors. This criterion ensures that the individual pulses are easily distinguishable in the burst and enables us to carry out a relatively ``clean" analysis, i.e., that the errors introduced in the fitting procedure are slight. 

We found that 21 GRBs from the \cite{li2021} catalogue match our complete list of criteria. All these bursts are collected up to June 2019. In addition, two bursts (GRB 110903, GRB 160625) that match our criteria appear in the GBM HEASARC database \citep{kienlin2020}.
Thus, our complete sample contains 22 different GRBs, showing a total of 61 pulses. These GRBs, along with their pulses, are listed in Table \ref{table: peer_ag} below. Two GRBs (GRB 171120(556), GRB 140329(295)) were excluded as they did not meet the $r^2> 0.7$ criteria. 

\subsection{Pulse Model Analysis} \label{sec:pulsemodel}

In order to analyse the pulses detected by the GBM instrument, we introduce a new fitting function different from the one used previously in the literature. 
Our new function is composed of two approximate sigmoid functions and is given in Equation  \ref{eq:peer_ag}. It can properly fit for the slopes, width and height of a pulse. 
It is capable of fitting for both FRED and symmetric pulse shapes as the parameters are flexible.

\begin{figure} 
    \includegraphics[width=0.3\textwidth]{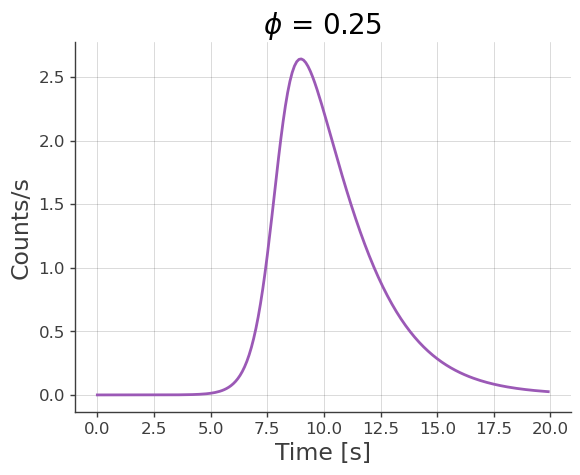} 
    \label{fig:PulseShape_0.25}
    \hfill
    \includegraphics[width=0.3\textwidth]{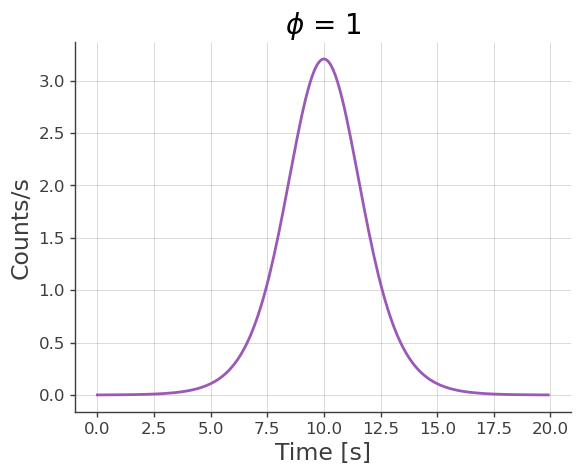} 
    \label{fig:PulseShape_1}
    \hfill
    \includegraphics[width=0.3\textwidth]{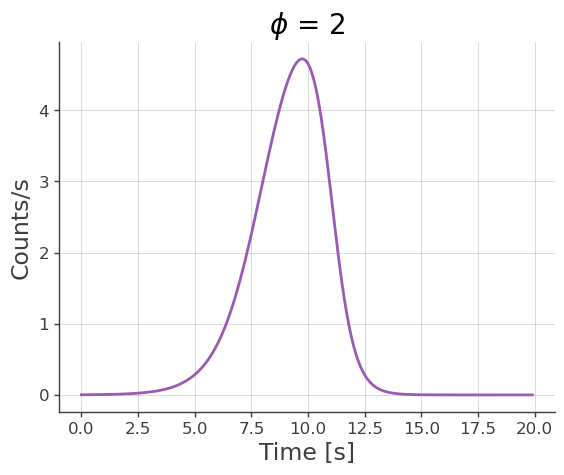}
    \label{fig:PulseShape_2}
\caption{ Examples of pulse shapes obtained by the fitting function given in Equation ~ \ref{eq:peer_ag}. Time is given in seconds. Here, Left:  $\phi$ = 0.25 (FRED-like). Middle: $\phi$ = 1 (Symmetric). Right:  $\phi$ = 2 (Symmetric-like). }
\label{fig:PulseShape}
 \end{figure}

\begin{figure}
    \includegraphics[width=0.3\textwidth]{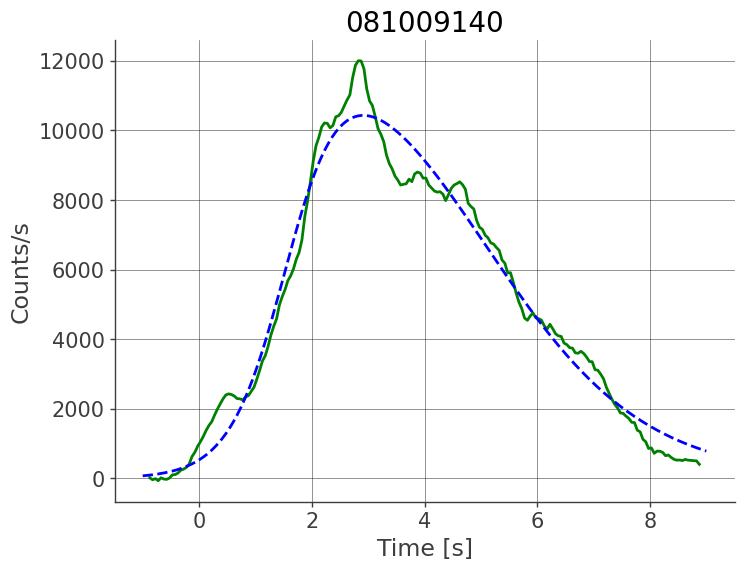} 
    \label{fig:fit_pg1}
    \hfill
    \includegraphics[width= 0.3\textwidth]{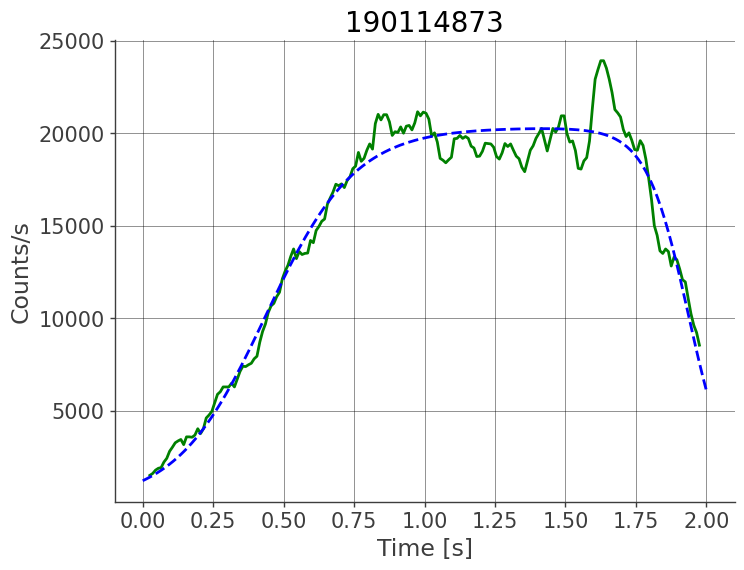}
    \label{fig:fit_pg2} \hfill
    \includegraphics[width = 0.3\textwidth]{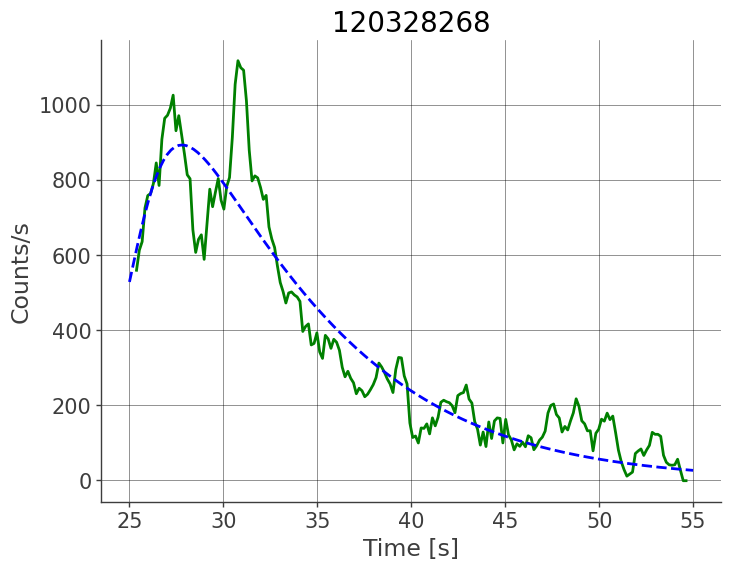}
    \label{fig:fit_pg3}
 \caption{Examples of fitting different pulses from the sample according to the pulse shape function. The solid green line represents the raw count-rate data from GBM, and the dashed blue line represents the pulse shape model fit according to Equation \ref{eq:peer_ag}. Left: The novel pulse function fit for the prompt light curve of GRB 081009 (bn 08109140). Here, the pulse shape parameter $\phi$ = 0.35, the pulse shape is \textbf{FRED-like} and it is a \textbf{single-peaked} pulse. Middle: The novel pulse function fit for the prompt light curve of GRB 190114 (bn 190114873). Here, the pulse shape parameter $\phi$ = 2.12, and the pulse shape is \textbf{Symmetric-like}. 
 Right: The novel pulse function fits the prompt light curve of GRB 120328 (bn 120328268). The pulse shape parameter $\phi$ = 0.17, and the pulse shape is FRED-like and it is a \textbf{combined-peak} pulse.
}
\label{fig:fit_eg}
\end{figure}

The empirical pulse model function is given by, 

\begin{equation}\label{eq:peer_ag}
     I(t) = \frac{A}{4} \times \left[{1-\tanh\left(\frac{1}{s_r}(t-r_r)\right)}\right] \times  \left[{1+\tanh\left(\frac{1}{s_l}(t-r_l)\right)} \right].
 \end{equation}
Here, \textit{I(t)} is given in units of count-rate,  $s_l$ and $s_r$ represent the rising (pre-peak) and decaying (post-peak) slopes of the light curve - smaller values imply steeper slopes (before and after the peak, respectively). The parameters $r_l$ and $r_r$ are the half-time of light curve rise and fall, respectively, and $A$ is a global normalization. The subscripts $l$ and $r$ indicate the left and right sides of the peak, respectively. This function, therefore, has a total of five degrees of freedom. It is designed for the best fits of a single peak pulse, enabling it to capture both symmetric and FRED-like behaviour on an equal footing. Examples of pulse shapes that can be obtained from fitting the function are given in Figure  \ref{fig:PulseShape}. Furthermore, this function enables good fits to pulses with elongated peaks (see examples in Figure \ref{fig:fit_eg}), often manifested by multiple light curve fluctuations.

{\bf Determining the pulse shapes.}
 By allowing for the rise and decay phase degree of freedom, we are able to capture the nuances in the curve shape rather than restricting it to be FRED-like or symmetric-like. 
We define the $\mathbf{pulse \ shape \ parameter \ (\phi) }$ as the ratio between the rising and decaying slopes, 
\begingroup

\begin{equation}
  \phi = \frac{s_l}{s_r}. \label{eq: shape}
\end{equation} 
\endgroup
This ratio indicates the asymmetry in the pulse shape. Examples of a few pulse shapes using Equation \ref{eq:peer_ag} are presented in Figure \ref{fig:PulseShape}.

{\bf Fitting procedure.}
The fits are carried using the $curvefit$ function of the python library  \texttt{scipy.optimize} \citep{virtanen2020}. The TTE data with $2\mu s$ resolution is binned during the source interval in such a way as to have up to 300 bins to account for the limitations of \texttt{scipy.optimize.curvefit()} function.
In practice, the light curve within a pulse is always somewhat fluctuating. Thus, one first has to identify the signal and discriminate from the noise. 
For that, after we fit the pulse using the function $I(t)$, defined in Equation \ref{eq:peer_ag} above, we then look at the residuals. Here the $r^2 > 0.7$ criterion ensures that the fits are acceptable.

The fit parameters are sensitive to the signal-to-noise ratio and spiky variations in the light curve. We acknowledge that the pulse selection criteria are somewhat arbitrarily chosen and, as such, have some degree of subjectivity; thus, one could potentially select two pulses as one due to marginal variations in the count-rate. Therefore, to be fully transparent, we have defined two types of pulses after fitting: single peak and combined peak pulses. If a given pulse has only one detectable peak, and the pulse takes several seconds to rise before and decay after the peak, then it is defined as a single-peak pulse. If, on the other hand, several spiky variations along with the detected pulse peak can be identified, these pulse fits are marked as combined peak pulses as defined below.


We assume that the subdominant peaks with a height that does not exceed 20\% of the peak count-rate of the pulse in 0.1s binning of the light curve are part of the dominant pulse emission. Any pulse with fluctuations that rise and decay in less than a second and with peak heights that exceed this 20\% level of peak count-rate is categorized as a combined peak pulse.
This categorisation of pulses enables us to be aware of the variations that we may have smoothed over while defining a pulse. Some pulses could have peak heights marginally over the bound and cannot be categorised into either type.
For example, in Figure \ref{fig:fit_eg}, the first subfigure of GRB 081009(140) is a single peak pulse, while the third one of GRB 130606(497) is a combined peak pulse. The second example of GRB 190114(873) represents a pulse which cannot be categorised into either type. 
We ensure that the signal-to-noise ratio is high by the criterion that at least one Bayesian bin of the pulse has $S > 20$. This condition, however, does not guarantee that all the bins, especially the ones at the beginning and at the end of each pulse, have the required S/N ratio. The allowed range of the count-rate at the beginning and at the end of the pulse, 48 - 52 \% ($50 \pm 2 ~ \%$) and the combined peak pulse classification takes into account these smaller subdominant variations and structures in the pulse envelope in our analysis. 
We have checked that this variation does not affect the obtained value of the rising or decaying slopes. Therefore, it does not affect our shape parameter, $\phi$.\\

Equation \ref{eq:peer_ag}, while similar to a logistic function, offers more flexibility than a simple Gaussian or logistic function, by allowing both the rise and decay slopes to be varied independently. This property also allows one to fit the pulses which have extended or plateau-like peaks, such as Figure \ref{fig:fit_pg2} [Right]. Although Equation \ref{eq:peer_ag} is a combination of two logistic functions, the resulting pulse shape $\phi$ parameter is more direct and more sensitive to the varied pulse shapes of GRBs than the traditional statistical measures like skewness or kurtosis. Moreover, Equation \ref{eq:peer_ag} has two main advantages over previous fit functions used in the literature \citep[e.g.,][]{Norris1996, norris2005}. The function used by \cite{norris2005} is an asymmetric function. 
As such, this function can not capture a pulse shape where the rise time exceeds the decay time,  whereas Equation (\ref{eq:peer_ag}) accommodates such cases with a pulse shape parameter $\phi > 1$. Furthermore, the fitting function used by \cite{norris2005}
characterizes the pulse asymmetries using a parameter $k$, where $k = 0$ represents symmetric pulses and values in the range $0 < k < 1$ correspond to asymmetric pulses with decay times longer than rise times. The limited range of $k$ values makes distinguishing between symmetric-like and FRED-like pulses more challenging. In contrast, in the model used here, distinguishing between a FRED-like pulse shape ($0 < \phi \le 0.3$) and symmetric-like pulses ($\phi> 1$) is more direct. We provide further details of the difference between the fitting functions in Appendix \ref{sec:Appendix}.


The burst name, pulse number, start and end times of each pulse, and the corresponding fitting parameters  A, $s_l$, $s_r$, $r_l$, $r_r$, the goodness of fit measurement $r^2$ after the fit to Equation \ref{eq:peer_ag}, along with the shape $\phi$ are given in Table \ref{table: peer_ag}.


\begin{deluxetable}{ccccccccccc}
\tablecaption{Pulse properties of the sample of 22GRBs (61 pulses) used in our study. Column 1:  Fermi-GBM trigger number (bn), Column 2: Pulse number (the order in which each pulse appears in a burst), Column~3: Start time of the pulse, Column 4: End time of the pulse, the novel pulse shape model fit parameters Column~5: A the normalisation, Column 6: $s_l$ the left side (rise) slope, Column 7: $s_r$ the right side (decay) slope, Column 8: $r_l$ half-time of rise, Column 9: $r_r$ half-time of decay, Column 10: $r^2$ the goodness of fit, Column 11: $\phi$ the pulse shape with error. }\label{table: peer_ag}
\tablehead{ 
\colhead{bn} & \colhead{Pulse} & \colhead{Pulse} & \colhead{Pulse}  & \colhead{A}  & \colhead{$s_l$} & \colhead{$s_r$} & \colhead{$r_l$} & \colhead{$r_r$} & \colhead{$r^2$} & \colhead{$\phi$ }  \\
\colhead{} & \colhead{Number} & \colhead{Start (s)} & \colhead{End (s)} & \colhead{}  & \colhead{} & \colhead{} & \colhead{}& \colhead{}& \colhead{}
}
\startdata
081009140  & 1        & -1.0    & 9.0     & 3409.02  & 0.99 & 2.84  & 1.58   & 5.04   & 0.97 & 0.35 $\pm$ 0.02         \\
081009140  & 2        & 33.0    & 53.0    & 8610.83  & 2.32 & 5.16  & 41.05  & 41.05  & 0.97 & 0.45 $\pm$ 0.03         \\
081215784  & 1        & 0.4   & 3.2   & 7235.82  & 0.13 & 0.77  & 1.33   & 1.43   & 0.82 & 0.17 $\pm$ 0.03         \\
081215784  & 2        & 4.2   & 9.0     & 4220.00  & 0.48 & 1.22  & 4.71   & 5.20   & 0.76 & 0.39  $\pm$ 0.10         \\
090618353  & 1        & -3.0    & 44.0    & 190.60   & 1.98 & 13.14 & 0.09   & 29.73  & 0.90 & 0.15 $\pm$ 0.01         \\
090618353  & 2        & 45.0    & 76.0    & 1338.52  & 5.61 & 3.32  & 60.50  & 73.46  & 0.92 & 1.69  $\pm$ 0.23         \\
090618353  & 3        & 76.0    & 103.0   & 12472.70 & 7.02 & 12.85 & 81.18  & 81.18  & 0.94 & 0.55  $\pm$ 0.09         \\
090618353  & 4        & 106.0   & 116.0   & 3092.71  & 3.13 & 21.18 & 108.28 & 220.28 & 0.94 & 0.15 $\pm$ 0.02         \\
091127976  & 1        & -0.4  & 1.1   & 27409.20 & 0.21 & 0.83  & 0.25   & 0.25   & 0.93 & 0.25  $\pm$ 0.06         \\
091127976  & 2        & 1.1   & 3.5   & 23355.80 & 0.18 & 0.69  & 1.24   & 1.24   & 0.87 & 0.25  $\pm$ 0.04         \\
091127976  & 3        & 5.9   & 12.0    & 11004.50 & 0.43 & 1.76  & 6.62   & 6.62   & 0.88 & 0.24  $\pm$ 0.03         \\
100719989 & 1        & 0.0     & 3.2   & 3730.00  & 0.78 & 0.61  & 2.50   & 2.69   & 0.91 & 1.29  $\pm$ 0.27         \\
100719989 & 2        & 3.2   & 9.0     & 3240.00  & 0.75 & 1.56  & 3.62   & 8.32   & 0.88 & 0.48 $\pm$ 0.10         \\
100719989 & 3        & 17.0    & 45.0    & 829.75   & 2.61 & 1.89  & 22.49  & 22.59  & 0.82 & 1.38  $\pm$ 0.70         \\
110301214 & 1        & -1.0    & 3.5   & 3180.46  & 1.24 & 0.69  & 1.25   & 3.33   & 0.92 & 1.79  $\pm$ 0.35         \\
110301214 & 2        & 3.5   & 11.5  & 28005.60 & 0.20 & 1.85  & 3.66   & 3.66   & 0.95 & 0.11  $\pm$ 0.02         \\
110625881 & 1        & 9.7   & 20.0    & 13450.40 & 0.23 & 2.46  & 10.41  & 10.41  & 0.89 & 0.09  $\pm$ 0.02         \\
110625881 & 2        & 20.0    & 26.0    & 2400.43  & 1.53 & 0.68  & 23.00  & 25.39  & 0.92 & 2.26  $\pm$ 0.26         \\
110903009 & 1        & -1.0    & 2.3   & 668.02   & 0.20 & 1.23  & -0.05  & 0.92   & 0.84 & 0.16  $\pm$ 0.02         \\
110903009 & 2        & 2.3   & 9.0     & 2365.66  & 0.55 & 2.37  & 3.68   & 3.90   & 0.93 & 0.23  $\pm$ 0.02         \\
110903009 & 3        & 19.9  & 28.0    & 451.95   & 0.26 & 3.48  & 20.28  & 23.01  & 0.81 & 0.07  $\pm$ 0.01         \\
120129580 & 1        & -0.7  & 2.5   & 12400.00 & 0.48 & 0.59  & 1.41   & 1.60   & 0.97 & 0.81  $\pm$ 0.15         \\
120129580 & 2        & 2.5   & 5.0     & 7460.00  & 0.36 & 0.69  & 2.87   & 2.97   & 0.96 & 0.52  $\pm$ 0.06         \\
120328268 & 1        & -1.0    & 16.0    & 7532.28  & 1.44 & 8.81  & 4.78   & 4.78   & 0.87 & 0.16  $\pm$ 0.03         \\
120328268 & 2        & 16.0    & 25.2  & 9022.00  & 2.38 & 4.53  & 19.51  & 19.51  & 0.85 & 0.53  $\pm$ 0.19         \\
120328268 & 3        & 25.0    & 55.0    & 2430.26  & 2.23 & 13.21 & 25.34  & 25.34  & 0.81 & 0.17 $\pm$ 0.05         \\
120711115 & 1        & 57.0    & 80.0    & 4450.60  & 3.84 & 14.04 & 66.20  & 66.20  & 0.85 & 0.27  $\pm$ 0.11         \\
120711115 & 2        & 80.0    & 120.0   & 437.30   & 9.73 & 5.74  & 84.73  & 104.94 & 0.83 & 1.70  $\pm$ 0.29         \\
120728434 & 1        & 5.0     & 40.0    & 4750.00  & 5.41 & 1.66 & 12.70  & 38.50  & 0.72 & 3.26 $\pm$ 0.93         \\
120728434 & 3        & 65.0    & 125.0   & 6741.96  & 0.68 & 20.42 & 68.27  & 68.27  & 0.86 & 0.03 $\pm$ 0.01         \\
130606497 & 1        & 0.0     & 12.0    & 7060.00  & 2.83 & 1.54  & 9.10   & 11.26  & 0.82 & 1.83 $\pm$ 0.54         \\
130606497 & 2        & 12.0    & 28.0    & 14827.50 & 1.08 & 7.30  & 12.30  & 12.30  & 0.95 & 0.15  $\pm$ 0.02         \\
130606497 & 4        & 48.0    & 75.0    & 8340.28  & 1.29 & 8.98  & 49.48  & 49.48  & 0.91 & 0.14  $\pm$ 0.02         \\
131014215 & 1        & 0.0     & 2.4   & 13100.00 & 0.60 & 0.43  & 1.70   & 2.09   & 0.94 & 1.39  $\pm$ 0.23         \\
131014215 & 2        & 2.4   & 5.5   & 3860.00  & 0.09 & 0.70  & 2.46   & 4.11   & 0.90 & 0.12  $\pm$ 0.02         \\
140213807 & 1        & -1.0    & 5.0     & 1155.91  & 0.77 & 4.37  & 0.56   & 2.59   & 0.91 & 0.18  $\pm$ 0.04         \\
140213807 & 2        & 5.0     & 17.0    & 2502.98  & 0.75 & 4.00  & 5.59   & 5.59   & 0.97 & 0.19  $\pm$ 0.02         \\
140508128 & 1        & -1.0    & 9.0     & 1812.89  & 1.17 & 1.10  & 4.00   & 6.06   & 0.72 & 1.06  $\pm$ 0.26         \\
140508128 & 2        & 23.0    & 29.0    & 3543.61  & 1.41 & 0.74  & 25.99  & 26.21  & 0.93 & 1.90  $\pm$ 0.29         \\
140508128 & 3        & 34.0    & 45.0    & 7791.07  & 0.31 & 2.14  & 38.80  & 38.80  & 0.76 & 0.15  $\pm$ 0.03         \\
150330828 & 1        & -1.0    & 3.8   & 338.16   & 2.06 & 0.92  & 2.50   & 3.08   & 0.80 & 2.23  $\pm$ 0.62         \\
150330828 & 2        & 3.8   & 11.5  & 5095.00  & 0.73 & 2.71  & 4.57   & 4.57   & 0.86 & 0.27  $\pm$ 0.03         \\
150330828 & 4        & 133.1 & 135.2 & 20695.50 & 0.36 & 1.87  & 133.23 & 133.23 & 0.79 & 0.19  $\pm$ 0.03         \\
150330828 & 5        & 135.1 & 142.7 & 13606.60 & 1.23 & 4.99  & 135.71 & 135.71 & 0.82 & 0.25  $\pm$ 0.05         \\
150330828 & 6        & 143.1 & 173.1 & 8648.92  & 1.52 & 7.92  & 145.09 & 145.09 & 0.92 & 0.19  $\pm$ 0.02         \\
\enddata
\end{deluxetable}

\begin{deluxetable}{ccccccccccc}
\setcounter{table}{1}
\renewcommand{\thetable}{}
\makeatletter
\renewcommand{\@makecaption}[2]{#2} 
\makeatother
\addtocounter{table}{-1}
\tablecaption{\textbf{Table  \ref{table: peer_ag} (Continued). }Pulse properties of the sample of 22 GRBs (61 pulses) used in our study.} 

\tablehead{ 
\colhead{bn} & \colhead{Pulse} & \colhead{Pulse} & \colhead{Pulse}  & \colhead{A}  & \colhead{$s_l$} & \colhead{$s_r$} & \colhead{$r_l$} & \colhead{$r_r$} & \colhead{$r^2$} & \colhead{$\phi$ }  \\
\colhead{} & \colhead{Number} & \colhead{Start (s)} & \colhead{End (s)} & \colhead{}  & \colhead{} & \colhead{} & \colhead{}& \colhead{}& \colhead{}
}
\startdata
151231443 & 1        & -2.0    & 20.0    & 5539.44  & 3.39 & 5.36  & 4.73   & 4.73   & 0.79 & 0.63  $\pm$ 0.25         \\
151231443 & 2        & 58.0    & 68.0    & 283.86   & 1.54 & 1.02  & 62.03  & 67.52  & 0.85 & 1.51  $\pm$ 0.21         \\
151231443 & 3        & 68.0    & 80.0    & 4575.32  & 2.47 & 3.45  & 69.81  & 69.81  & 0.79 & 0.72  $\pm$ 0.29         \\
160625945 & 1        & -1.0    & 1.5   & 997.77   & 0.20 & 0.28  & 0.03   & 0.70   & 0.88 & 0.71  $\pm$ 0.09         \\
160625945 & 2        & 184.0   & 192.0   & 4690.11  & 0.64 & 1.64  & 188.00 & 190.93 & 0.96 & 0.39  $\pm$ 0.05         \\
160625945 & 3        & 192.0   & 197.0   & 6144.43  & 4.24 & 1.92  & 194.50 & 196.51 & 0.86 & 2.21 $\pm$ 0.48         \\
160625945 & 4        & 198.0   & 235.5 & 7516.76  & 1.51 & 8.46  & 198.74 & 198.74 & 0.96 & 0.18 $\pm$ 0.02         \\
160802259 & 1        & -1.0    & 3.5   & 1953.18  & 0.30 & 1.05  & 0.20   & 2.78   & 0.92 & 0.28  $\pm$ 0.03         \\
160802259 & 2        & 3.5   & 5.0     & 1260.00 & 0.18 & 1.11  & 3.64   & 3.64   & 0.72 & 0.16 $\pm$ 0.05         \\
160802259 & 3        & 4.9   & 9.9   & 13570.10 & 0.21 & 0.80  & 5.19   & 5.19   & 0.94 & 0.26  $\pm$ 0.03         \\
160802259 & 4        & 15.0    & 21.0    & 18759.10 & 0.23 & 1.40  & 15.72  & 15.72  & 0.97 & 0.16 $\pm$ 0.01         \\
171227000 & 1        & -1.0    & 24.0    & 34000.00 & 3.58 & 3.24  & 19.00  & 20.56  & 0.93 & 1.10  $\pm$ 0.18         \\
171227000 & 2        & 24.0    & 35.0    & 11600.00 & 1.97 & 3.93  & 27.00  & 27.00  & 0.71 & 0.50  $\pm$ 0.23         \\
171227000 & 3        & 35.0    & 63.0    & 5520.00   & 2.28 & 12.90 & 36.20  & 42.55  & 0.73 & 0.18  $\pm$ 0.05         \\
190114873 & 1        & 0.0     & 2.0     & 5080.00  & 0.32 & 0.15  & 0.44   & 1.50   & 0.94 & 2.12  $\pm$ 0.22         \\
190114873 & 2        & 2.0     & 15.0    & 4580.00  & 0.19 & 0.51  & 2.30   & 3.62   & 0.89 & 0.37  $\pm$ 0.11  \\
\enddata

\end{deluxetable}

\subsection{Spectral analysis} \label{sec:2.4}

In addition to fitting the light curves, we also fitted the spectra of the various pulses to identify any correlations.
The spectral analysis is done using the Bayesian spectral analysis package 3ML  \citep{Vianello2015}. 
We follow the method of the {\it Fermi}-GBM GRB time-integrated \citep{Goldstein2012,Gruber2014} and time-resolved catalogues \citep{Yu2016,YDR19}. We select the brightest one to three NaI detectors and a BGO detector. The source interval is marked from the beginning of a rise phase to the end of the decay phase from the brightest NaI detector light curve after an initial 0.1s binning. 
The background interval is marked a few 10s of seconds before and after the source interval. In some cases where the pulse shapes were separated by long quiescent intervals, we included three background intervals. A polynomial fit is applied to the total photon count in each of the 128 energy channels of the TTE (time-tagged data) of all the selected detectors. The order of the polynomial fit is obtained after a likelihood ratio test. This polynomial fit is used to extract the background count rate during the source interval.

The light curves are binned using Bayesian blocks \citep{Scargle2013} to carry a time-resolved spectral analysis. The Bayesian block binning method identifies the intervals with a constant Poisson rate, and the data is rebinned in such a way as to minimize the variations in emission. We use the common false positive probability $p_0 = 0.01$ \citep[e.g.,][]{Vianello2018b, burgess2019, YDR19}  to the brightest NaI detector TTE light curve and this binning is applied to the other detectors. We note our implicit assumption that a small variation in the light curve corresponds to a small spectral variation. 

Adopting statistical significance $S$, it integrates the information about the signal-to-noise ratio for the Poisson sources with Gaussian backgrounds \citep[see][for the definition of $S$]{Vianello2018b}. The spectral parameters are typically well-constrained for bins with
statistical significance $S \ge 20$. Hence, for pulses with duration $\le 5$~s, we require selected pulses to have at least one Bayesian block with $S \ge 20$ and for pulses with duration $\ge 5$~s two Bayesian blocks with $S \ge 20$. 

The pulse spectra are fitted with the empirical ``Band" model \citep{Band1993} to keep the analysis consistent, as most of the pulses from \citet{li2021} obtained better fits with the Band model (compared to the ``cut-off power law (CPL)" model). 
Each pulse is split into a different number of time bins, and the low energy spectral index $\alpha$ can, in principle, vary during the pulse duration. For each physical emission model (e.g., synchrotron, thermal, etc.), there are limits on the maximal (highest) value of $\alpha$ that is theoretically allowed. If, during a pulse, any bin's $\alpha$ value violates such a limit, the emission model is rejected. This method was introduced in \citet{YDR19, Acuner2019, DPR20}. We point out that an underlying assumption is that a pulse results from only a single emission mechanism. Hence, we choose one bin containing the highest value of the low energy spectral index, denoted by $\alpha_{\max}$, for characterising the emission mechanism responsible for the pulse. The corresponding peak energy $E_{pk,\max}$ [keV] of the bin along with $\alpha_{\max}$ are given in Table \ref{table: spectral_prop}.

\begin{deluxetable}{cccccccccccc}
\setcounter{table}{2}
\addtocounter{table}{-1}  
\tablecaption{Spectral properties of the sample of 22 GRBs (61 pulses) used in our study. Column 1: Fermi-GBM trigger number, Column 2: Pulse number, Column 3: Burst duration $T_{90}$, Column 4: maximum low-energy spectral index $\alpha_{\max}$, Column~5: Corresponding peak energy $E_{pk, \max}$. }  
\tablehead{ 
\colhead{bn} & \colhead{Pulse} & \colhead{$T_{90}$} & \colhead{$\alpha_{\max}$}  & \colhead{$E_{pk, \max}$}  \\
\colhead{} & \colhead{Number} & \colhead{(s)} & \colhead{} & \colhead{(keV)}  
}\label{table: spectral_prop}
\startdata
081009140  & 1     & 54.0  & $-0.48^{+0.13}_{-0.13}$       & $39.80^{+1.19}_{-0.88}$       \\
081009140  & 2     & 54.0  & $-0.24^{+0.71}_{-0.52}$      & $19.95^{+2.22}_{-2.40}$        \\
081215784  & 1     & 8.6   & $-0.30^{+0.04}_{-0.04}$  & $769.07^{+44.60}_{-45.90}$      \\
081215784  & 2     & 8.6   & $-0.26^{+0.04}_{-0.04}$    & $326.03^{+32.20}_{-27.30}$     \\
090618353  & 1     & 119.0 & $-0.66^{+0.07}_{-0.06}$    & $192.43^{+12.74}_{-15.68}$    \\
090618353  & 2     & 119.0 & $-0.78^{+0.11}_{-0.11}$     & $220.83^{+31.23}_{-32.55}$  \\
090618353  & 3     & 119.0 & $-1.02^{+0.08}_{-0.07}$    & $108.09^{+7.28}_{-6.99}$   \\
090618353  & 4     & 119.0 & $-1.16^{+0.18}_{-0.18}$     & $73.86^{+10.00}_{-8.06}$   \\
091127976  & 1     & 12.4  & $-0.10^{+0.21}_{-0.17}$    & $47.63^{+3.26}_{-2.69}$     \\
091127976  & 2     & 12.4  & $-1.23^{+0.06}_{-0.07}$    & $277.01^{+57.22}_{-35.38}$  \\
091127976  & 3     & 12.4  & $-0.56^{+0.43}_{-0.48}$    & $12.43^{+1.47}_{-1.33}$    \\
100719989 & 1     & 45.0  & $-0.23^{+0.10}_{-0.09}$     & $460.55^{+45.70}_{-39.60}$      \\
100719989 & 2     & 45.0  & $-0.21^{+0.15}_{-0.13}$      & $261.14^{+32.20}_{-27.60}$      \\
100719989 & 3     & 45.0  & $-0.74^{+0.10}_{-0.10}$      & $185.51^{+25.10}_{-23.10}$      \\
110301214 & 1     & 12.5  & $-0.41^{+0.10}_{-0.11}$     & $122.637^{+7.54}_{-7.09}$     \\
110301214 & 2     & 12.5  & $-0.70^{+0.06}_{-0.05}$     & $121.50^{+6.03}_{-4.96}$     \\
110625881 & 1     & 16.3  & $-0.20^{+0.10}_{-0.11}$      & $157.14^{+13.10}_{-9.81}$     \\
110625881 & 2     & 16.3  & $-0.48^{+0.10}_{-0.07}$     & $113.12^{+7.28}_{-7.20}$       \\
110903009 & 1     & 29.0  & $-0.90^{+0.15}_{-0.11}$      & $50.02^{+2.57}_{-2.11}$ \\
110903009 & 2     & 29.0  & $-0.98^{+0.09}_{-0.09}$      & $34.70^{+0.95}_{-1.05}$     \\
110903009 & 3     & 29.0  & $-1.12^{+0.10}_{-0.11}$      & $128.4^{+29.6}_{-27.5}$       \\
120129580 & 1     & 5.7   & $0.62^{+0.44}_{-0.47}$       & $56.17^{+6.28}_{-8.17}$       \\
120129580 & 2     & 5.7   & $0.34^{+0.79}_{-0.85}$      & $43.65^{+9.37}_{-8.27}$       \\
120328268 & 1     & 56.0  & $-0.44^{+0.04}_{-0.04}$    & $226.72^{+11.00}_{-10.60}$        \\
120328268 & 2     & 56.0  & $-0.56^{+0.05}_{-0.05}$    & $195.80^{+14.00}_{-12.90}$         \\
120328268 & 3     & 56.0  & $-1.05^{+0.08}_{-0.09}$    & $114.69^{+12.70}_{-13.10}$      \\
120711115 & 1     & 63.0  & $-0.89^{+0.04}_{-0.03}$     & $1564.54^{+155.00}_{-169.00}$      \\
120711115 & 2     & 63.0  & $-0.65^{+0.12}_{-0.13}$      & $1033.21^{+303.00}_{-304.00}$       \\
120728434 & 1     & 120.0 & $-0.19^{+0.13}_{-0.13}$      & $97.30^{+8.32}_{-7.07}$        \\
120728434 & 3     & 120.0 & $0.10^{+0.15}_{-0.16}$        & $64.81^{+3.17}_{-3.73}$      \\
130606497 & 1     & 75.0  & $-0.49^{+0.15}_{-0.23}$     & $244.52^{+80.40}_{-58.50}$      \\
130606497 & 2     & 75.0  & $-0.87^{+0.04}_{-0.04}$    & $501.20^{+58.50}_{-59.50}$       \\
130606497 & 4     & 75.0  & $-0.58^{+0.05}_{-0.05}$   & $237.31^{+13.00}_{-16.00}$          \\
131014215 & 1     & 5.5   & $-0.02^{+0.07}_{-0.08}$   & $413.14^{+31.20}_{-24.40}$      \\
131014215 & 2     & 5.5   & $0.04^{+0.10}_{-0.12}$    & $158.16^{+9.88}_{-8.90}$       \\
140213807 & 1     & 18.0  & $-0.7^{+0.09}_{-0.09}$      & $108.88^{+9.10}_{-9.51}$       \\
140213807 & 2     & 18.0  & $-0.50^{+0.20}_{-0.21}$       & $61.80^{+5.85}_{-6.15}$        \\
140508128 & 1     & 58.0  & $-0.47^{+0.06}_{-0.06}$    & $391.41^{+24.70}_{-29.00}$        \\
140508128 & 2     & 58.0  & $-0.45^{+0.06}_{-0.06}$    & $195.14^{+8.87}_{-9.37}$    \\
140508128 & 3     & 58.0  & $-1.04^{+0.08}_{-0.09}$   & $185.89^{+26.80}_{-28.40}$      \\
150330828 & 1     & 174.1 & $-0.12^{+0.09}_{-0.09}$      & $379.41^{+32.70}_{-28.40}$     \\
150330828 & 2     & 174.1 & $-0.05^{+0.17}_{-0.21}$      & $149.12^{+15.40}_{-15.20}$      \\
150330828 & 4     & 174.1 & $-0.66^{+0.04}_{-0.04}$    & $245.60^{+13.30}_{-16.40}$       \\
150330828 & 5     & 174.1 & $-0.66^{+0.05}_{-0.05}$     & $183.40^{+11.70}_{-13.30}$     \\
150330828 & 6     & 174.1 & $-0.75^{+0.04}_{-0.05}$   & $200.00^{+11.80}_{-14.20}$         
\enddata
\end{deluxetable}

\begin{deluxetable}{cccccccccccc}
\addtocounter{table}{-1}
\renewcommand{\thetable}{}
\makeatletter
\renewcommand{\@makecaption}[2]{#2} 
\makeatother
\tablecaption{\textbf{Table \ref{table: spectral_prop} (Continued). }Spectral properties of the sample of 22 GRBs (61 pulses) used in our study. }
\tablehead{ 
\colhead{bn} & \colhead{Pulse} & \colhead{$T_{90}$} & \colhead{$\alpha_{\max}$}  & \colhead{$E_{pk, \max}$}  \\
\colhead{} & \colhead{Number} & \colhead{(s)} & \colhead{} & \colhead{(keV)}  
}
\startdata
151231443 & 1     & 82.0  & $-0.66^{+0.10}_{-0.11}$    & $252.34^{+22.70}_{-24.00}$        \\
151231443 & 2     & 82.0  & $-0.73^{+0.21}_{-0.19}$    & $146.25^{+22.60}_{-21.60}$      \\
151231443 & 3     & 82.0  & $-0.51^{+0.11}_{-0.10}$     & $144.68^{+8.99}_{-7.72}$     \\
160625945 & 1     & 236.5 & $-0.13^{+0.13}_{-0.14}$     & $68.74^{+2.51}_{-2.64}$       \\
160625945 & 2     & 236.5 & $-0.64^{+0.02}_{-0.02}$ & $741.80^{+33.80}_{-35.80}$       \\
160625945 & 3     & 236.5 & $-0.71^{+0.02}_{-0.02}$    & $608.43^{+26.50}_{-23.80}$      \\
160625945 & 4     & 236.5 & $-0.68^{+0.02}_{-0.02}$    & $767.11^{+32.70}_{-29.50}$      \\
160802259 & 1     & 22.0  & $-0.08^{+0.09}_{-0.09}$    & $338.95^{+23.60}_{-27.60}$     \\
160802259 & 2     & 22.0  & $-0.43^{+0.15}_{-0.16}$      & $158.52^{+20.80}_{-20.30}$     \\
160802259 & 3     & 22.0  & $-0.49^{+0.08}_{-0.06}$    & $263.93^{+14.30}_{-19.40}$      \\
160802259 & 4     & 22.0  & $-0.09^{+0.14}_{-0.14}$     & $124.65^{+10.60}_{-9.47}$      \\
171227000 & 1     & 61.0  & $-0.46^{+0.06}_{-0.07}$   & $1072.20^{+114.00}_{-120.00}$        \\
171227000 & 2     & 61.0  & $-0.66^{+0.06}_{-0.06}$     & $1148.45^{+149.00}_{-157.00}$     \\
171227000 & 3     & 61.0  & $-0.96^{+0.10}_{-0.10}$     & $201.76^{+40.83}_{-37.77}$   \\
190114873 & 1     & 170.0 & $-0.39^{+0.03}_{-0.03}$      & $811.60^{+29.73}_{-31.22}$    \\
190114873 & 2     & 170.0 & $-0.08^{+0.02}_{-0.03}$    & $840.60^{+24.61}_{-19.54}$    
\enddata
\end{deluxetable}

\section{Results}
\subsection{Individual correlations} \label{sec:obsrvations} 
We studied the pulse shapes and multiple correlations between the parameters. Through these observations, we investigate the emission origins of GRBs.
A summary of the basic properties of each of the 61 pulses obtained from the 22 GRBs is given in Table \ref{table: peer_ag}.
 
\subsection{The pulse shape studies} \label{sub: pulse shape studies}

\begin{figure}[ht!]
    \centering
    \includegraphics[width=15cm]{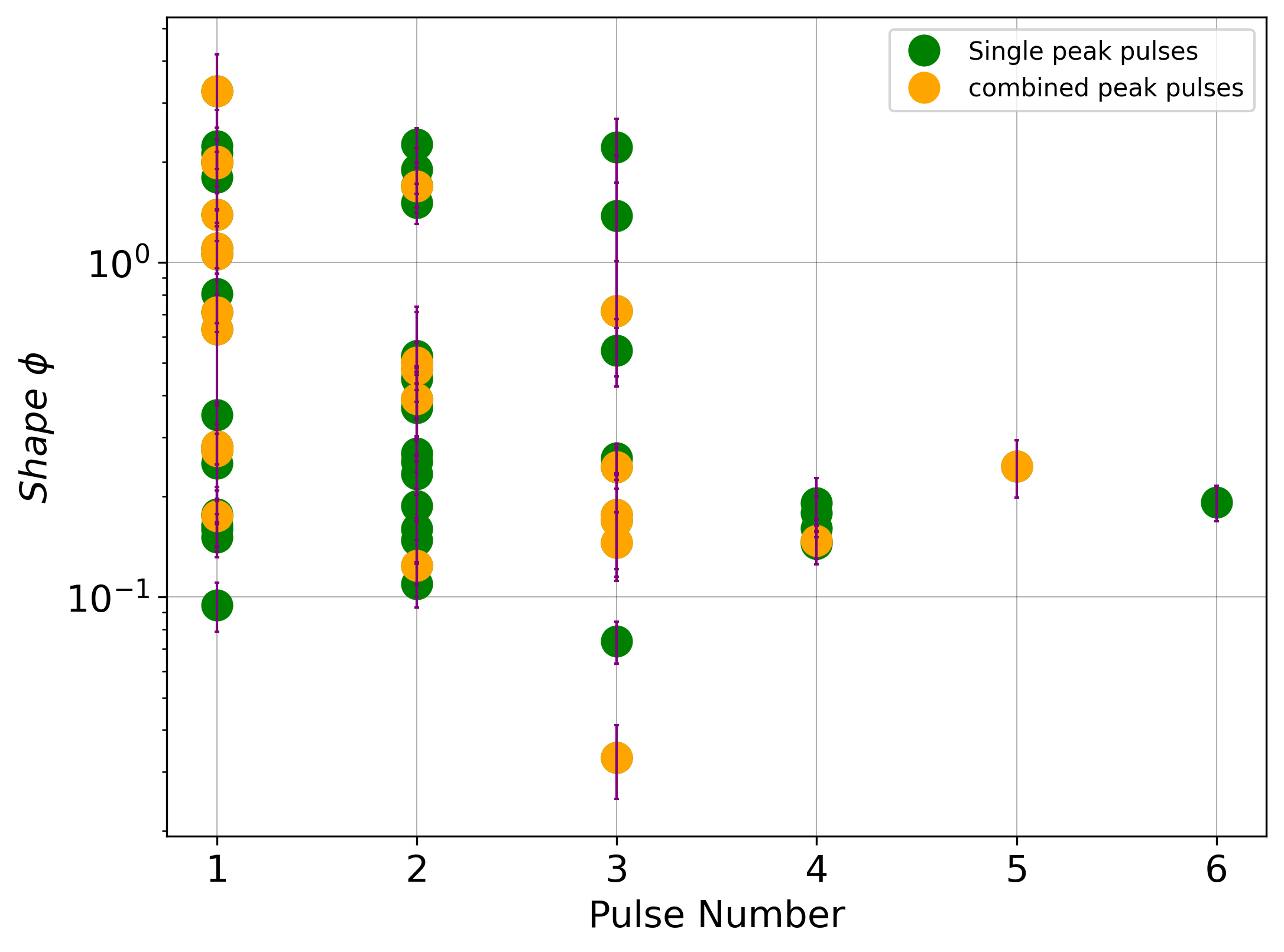}
    \caption{The shape versus pulse number relation. The green and orange dots present the single and combined pulses, respectively. Errors are in one $\sigma$ significance.}
    \label{fig:shape_plsno_err}
\end{figure}

In Figure \ref{fig:shape_plsno_err}, we plot the shape of each of the 61 pulses in our sample as defined in Equation \ref{eq: shape} above, as a function of the pulse number, i.e., the order in which the pulse appears within a burst (1$^{st}$, 2$^{nd}$, 3$^{rd}$, etc.). 
Pulse shape values lower than 0.3 can be identified as FRED-like shapes, and values higher than 1 are symmetric-like shapes. We further discriminate between single peak pulses and combined peak pulses as explained in section \ref{sec:pulsemodel}.

This analysis yields three key findings. Firstly, the shape function values span a broad range from 0 to 3. This range suggests a diversity in pulse shapes, varying from symmetric-like to FRED-like. Specifically, we can categorize the regions as follows: 0 - 0.3 represents pure FRED-like shapes, 1 - 3 corresponds to symmetric-like shapes, and the intermediate values indicate a mixed zone. The unexpected wide variation in pulse shapes suggests different physical origins, potentially varying energy dissipation radii, radiative mechanisms, or a combination of both.

Secondly, our findings show that approximately $\sim$ 26\% (16 out of 61) of the pulses exhibit a symmetric-like shape. Additionally, $\sim$ 51\% (31 out of 61) of the pulses fall under FRED-like shapes. When analyzing the first pulses, $\sim$ 41\% (9 out of 22) are symmetric-like. This percentage decreases to $\sim$ 24\% (5 out of 21) for second pulses and further drops to $\sim$ 18\% (2 out of 11) for third pulses.

\begin{figure}[ht!]
    \centering
    \includegraphics[width=12cm]{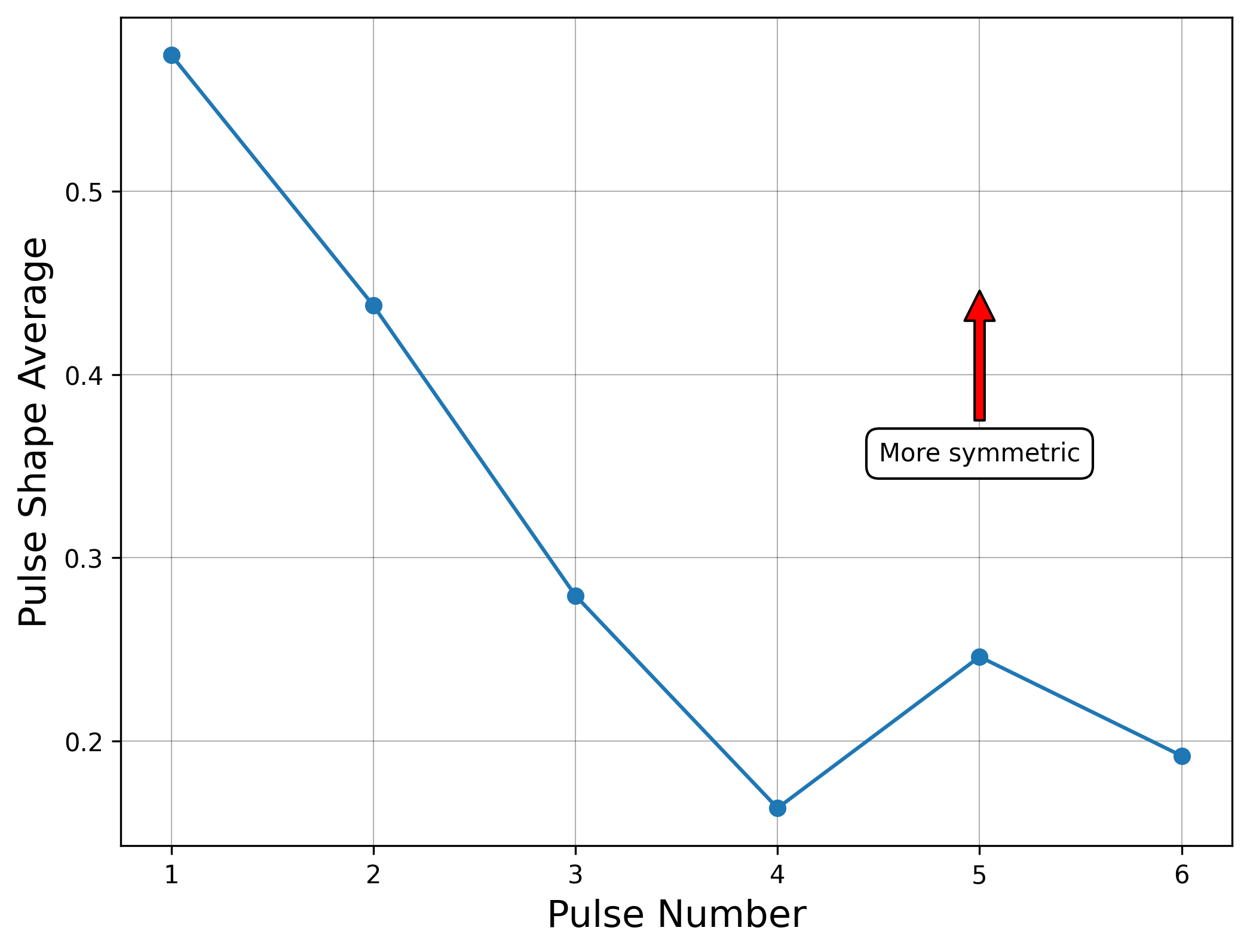}
    \caption{The average pulse shape values of each pulse number are indicated. }
    \label{fig:plsavg}
\end{figure}

Thirdly, the shape parameter steadily decreases with the increase of pulse number: later pulses tend to have much more FRED-like shapes than early pulses in a given GRB. This is shown in Figure \ref{fig:plsavg}, where we plot the geometric mean of pulse shape values corresponding to each pulse number. 

When we applied the correlation analysis between pulse shape and pulse number, we found that Spearman's rank correlation coefficient $r_s$ parameter for all the pulses is -0.34, with p-value = 0.008. This indicates a mild anti-correlation.
As time evolves, the shape of the pulses changes from symmetric-like to FRED-like. The initial pulses in our bursts tend to be more symmetric-like; this result contradicts some previous claims \citep{Norris1996} arguing for a FRED-like dominance in light curves. This result suggests that the radiative mechanism at different radii is likely changing throughout the burst.

It is difficult to explain this result within the context of the synchrotron (optically thin) emission model, in which the energy dissipation, which is the source of the observed photons, occurs far above the photosphere. This is because, in such a scenario, the rise time and decay time of a pulse are expected to have a different physical origin; there is no apriori reason to expect that both rise and decay phases would occur on similar time scales. However, such a result is more naturally explained if the dissipation occurs close to the photosphere due to the diffusion of photons in space and time (see further discussion below).

\begin{figure} 
    \includegraphics[width=0.49\textwidth]{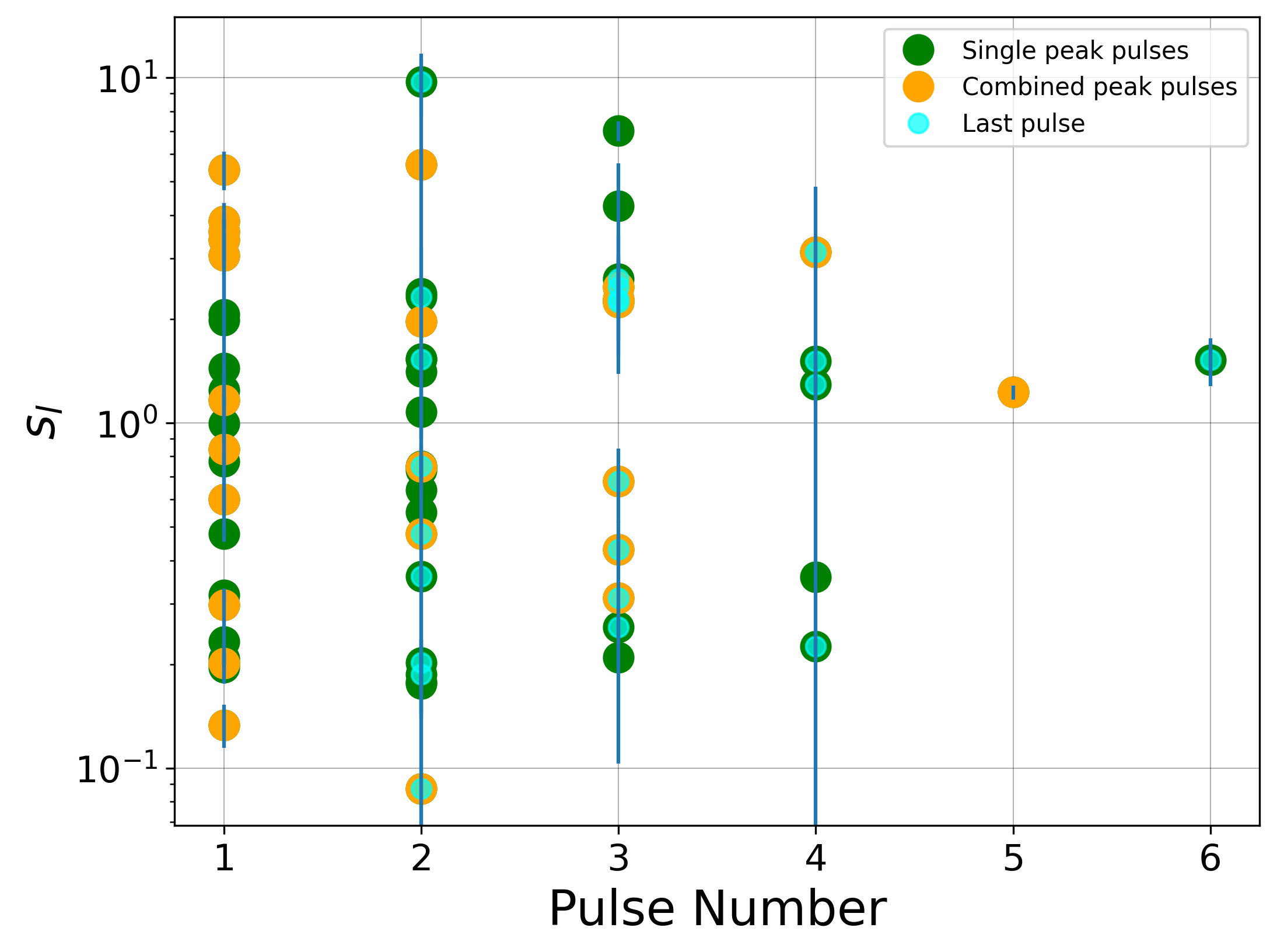} 
    \hfill
    \includegraphics[width=0.49\textwidth]{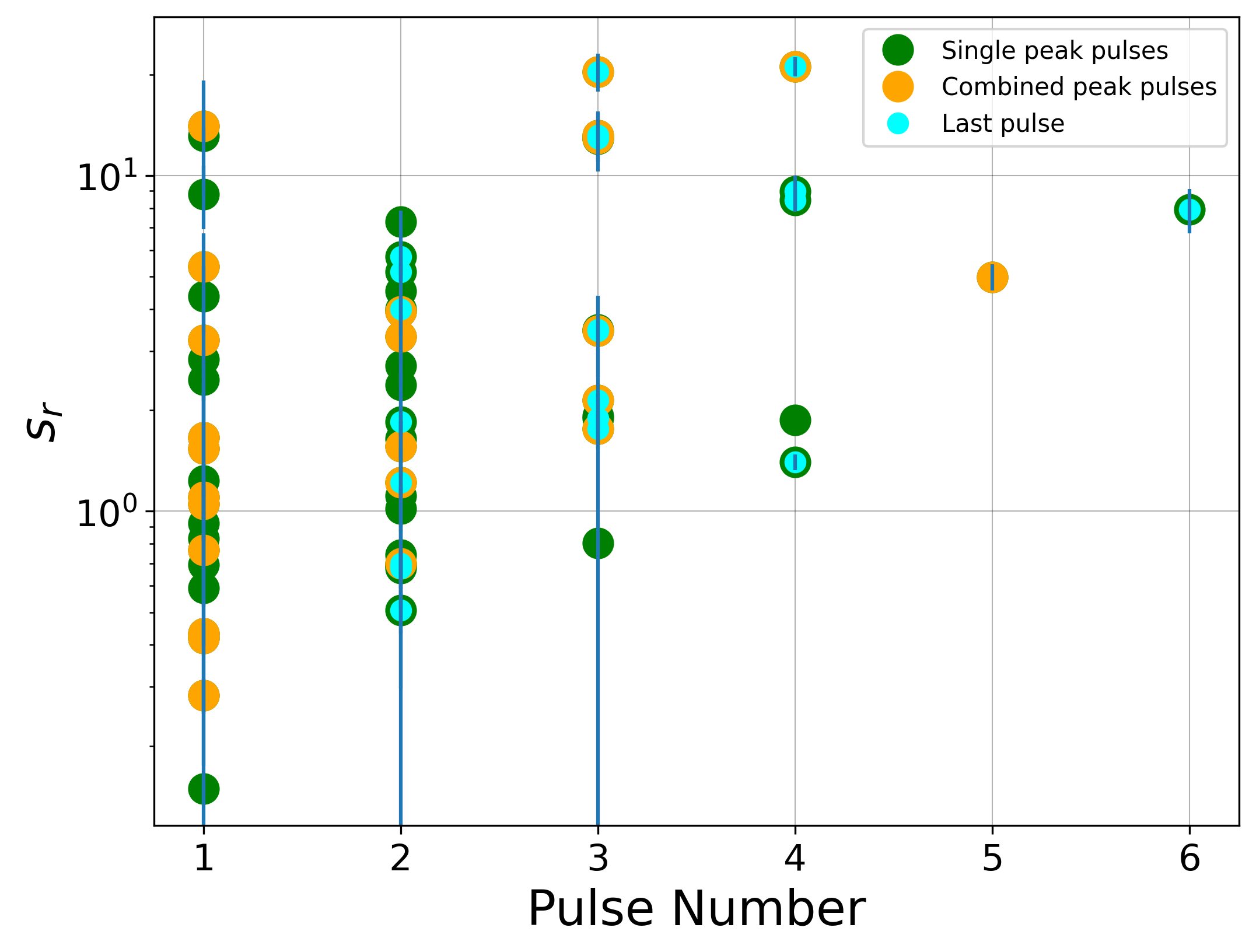} 
    \caption{Left: $s_l$ (rising slope) - Pulse number relation. The green and orange dots represent single and combined peak pulses, respectively. The last pulses of each GRBs are also marked here. Right: $s_r$ (decaying slope)- Pulse number relation. The green and orange dots represent single and combined peak pulses respectively. The last pulses of each GRBs are also marked here.} \label{fig: slvsr}
\end{figure}

There is a significant difference in Spearman's correlations between single peak and combined peak pulse populations. The $r_s$ value for the single peaked pulses is $-0.10$ with a p-value = 0.56. However, the $r_s$ value for the combined peak pulse is $-0.62$ with a p-value = 0.001, indicating a stronger anti-correlation. 
In order to understand the origin of this, we plot 
in Figure \ref{fig: slvsr} the rise and decay slopes, $s_l$ and $s_r$ as functions of the pulse number. Spearman correlation analysis between $s_l$ and pulse number shows that the $r_s$ parameter is 0.08 with a p-value = 0.53. In the case of the single peak pulses, the $r_s$ parameter is 0.14 with p-value = 0.40, and for combined peak pulses, the $r_s$ parameter is $-0.02$ with p-value = 0.91. Additionally, the correlation was applied to the last pulses of each GRB and the resultant $r_s$ parameter is 0.21 with p-value = 0.34.
This leads to the conclusion that the $s_l$ parameter does not change with pulse number. 

However, the same analysis on $s_r$ and pulse number shows a stronger correlation for all pulses, given by $r_s$ parameter 0.39, p-value = 0.001. For the single peak pulses, the $r_s$ parameter is 0.25 with p-value = 0.13, and for the combined peak pulses, the $r_s$ parameter is 0.58 with p-value = 0.003. This strong correlation in the case of the $s_r$ parameter with pulse number for combined peak pulses is reflected in the strong correlation for shape and pulse number for combined peak pulses. When the correlation is carried out for only the last pulses of each GRB the $r_s$ parameters become 0.56 with p-value = 0.006.
The stronger correlation of $s_r$ indicates that later pulses tend to have a longer decay phase, implying a FRED-like behaviour of the pulses that increases with pulse number, particularly at the last pulses. There can be several explanations for this behaviour. One possibility is a lower Lorentz factor at later pulses, resulting in a longer decay time scale due to high latitude emission. 



\subsection{Relation between the spectral shape and pulse shape} \label{sub: spectral shape studies}

\begin{figure}[ht!]
\centering
    \includegraphics[width = 14cm]{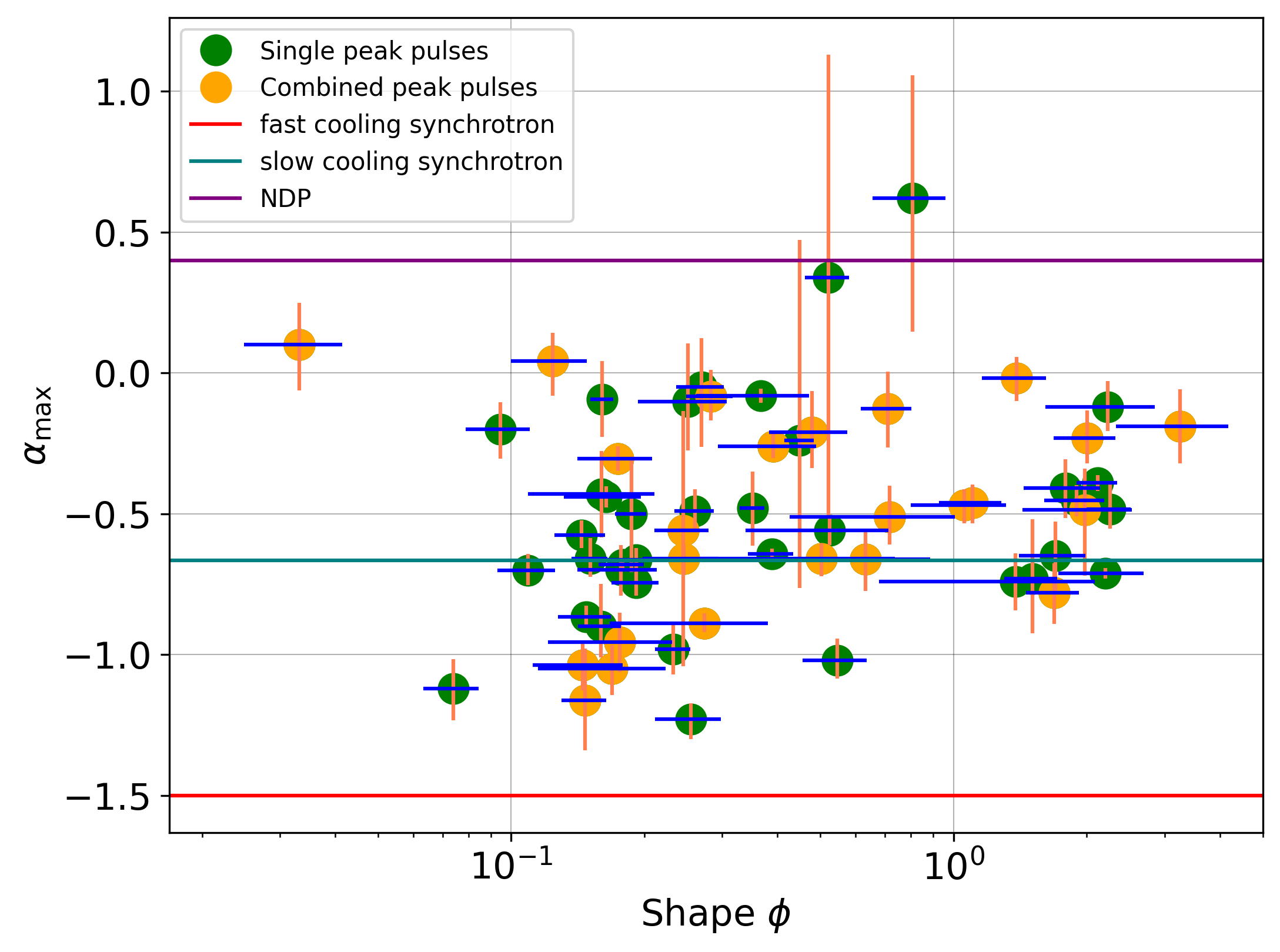}
    \caption{The $\alpha_{\max}$ versus shape relation. The green and orange data points are single-peak pulses and combined-peak pulses, as in Figure \ref{fig:shape_plsno_err}. The red, green and purple lines represent the upper limits for the fast and slow cooling synchrotron and the theoretical non-dissipative photospheric (NDP) emission, respectively.}
    \label{fig:alpha_shape_emi}
\end{figure}

We next examined the possible correlation between the pulse shape and spectral shape. Our main motivation is to examine the possibility raised above of different emission mechanisms for explaining symmetric-like and FRED-like pulses. 
We show in Figure \ref{fig:alpha_shape_emi} the highest value of the low energy spectral index, $\alpha_{\max}$, as a function of the pulse shape for the 61 pulses in our sample. 
The value of $\alpha_{\max}$ of each individual pulse is obtained following the method discussed in section \ref{sec:2.4} above. 

Correlation analysis shows that Spearman's, when applied to the full sample, is
$r_s = 0.23$, with p-value = $0.07$.
This indicates a trend between the maximal value of the low energy spectral index, $\alpha_{\max}$, and the pulse shapes: the more FRED-like pulses tend to have softer $\alpha_{\max}$ values. This may suggest a non-thermal emission origin. Generally, one finds that the more symmetric-like pulses have steeper spectral slopes (larger values of $\alpha_{\max}$), which can be interpreted as having a thermal (or quasi-thermal) origin. 

The lower $\alpha_{\max}$ values of FRED-like pulses may point to the idea that they originate at a large radius in the jet; hence, their spectra appear non-thermal, and the asymmetry in the light curve is due to light aberration. On the other hand, the higher value of $\alpha_{\max}$ of symmetric-like pulses hints towards the possibility that they originate at smaller radii, where the photospheric component is more pronounced, and photon diffusion below the photosphere symmetrizes the pulse.
There is also no substantial difference between the single peak and combined peak pulse results, as we can see from Spearman's 
$r_s$ values of $0.24$ and $0.20$ respectively. 

\begin{figure}[ht!]
\centering
    \includegraphics[width = 12cm]{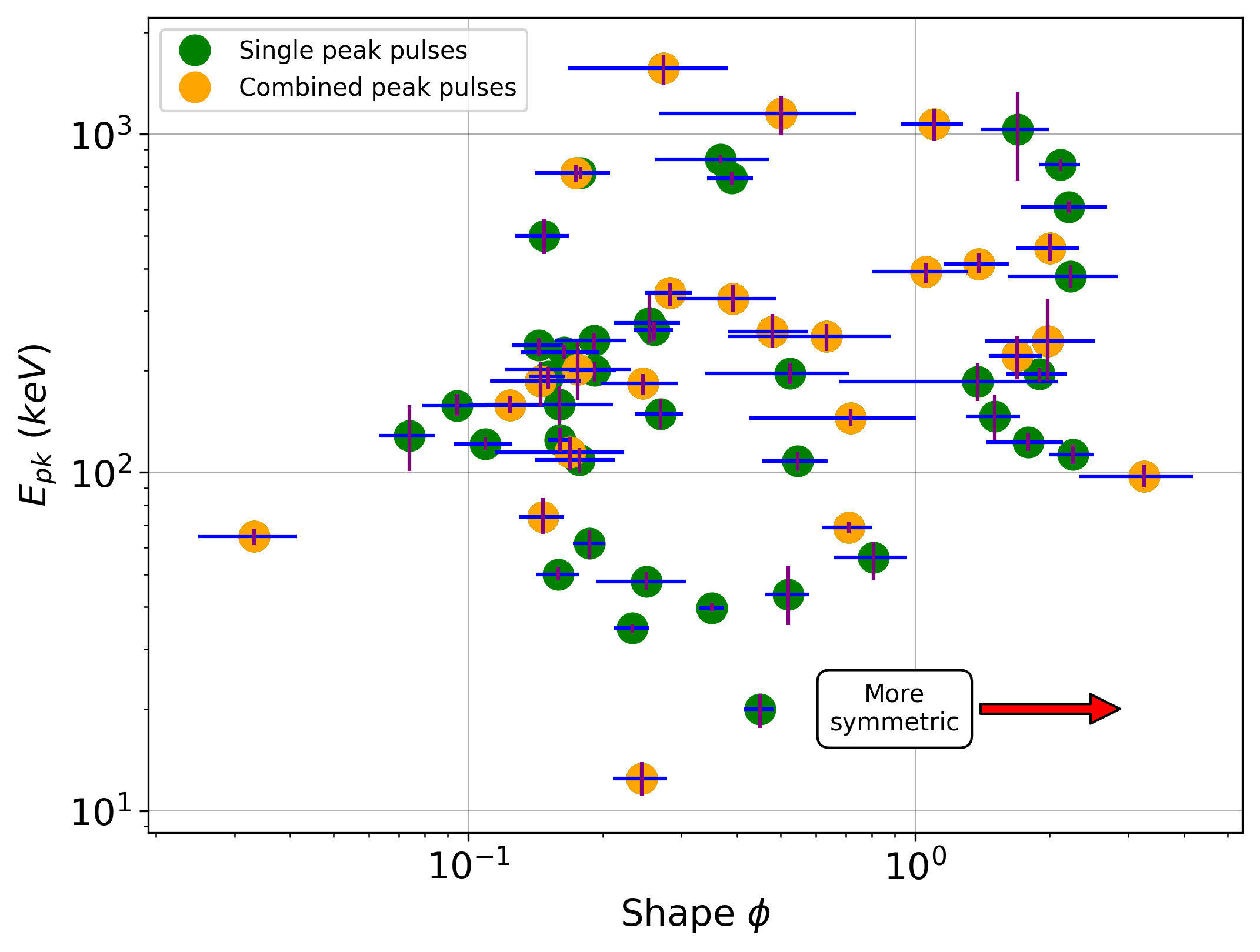}
    \caption{The $E_{pk}$ versus shape relation. The green and orange data points are single peak pulses and combined peak pulses, respectively. Errors are in one $\sigma$ significance.}
    \label{fig: epk_shape}
\end{figure}
The spectral index $\alpha_{\max}$ of all the pulses presented in Figure \ref{fig:alpha_shape_emi} is above the theoretical prediction of the spectral index in the ``fast cooling" synchrotron emission model (red line in Figure \ref{fig:alpha_shape_emi}). This indicates that the $\alpha_{\max}$ values are incompatible with the efficient cooling of electrons in the jet. One pulse (GRB 120129/bn120129580) is above the NDP line (purple line), indicating a likely photospheric origin. But 67\% of the pulses are between the NDP line and the slow cooling synchrotron line, showing a bias towards thermal overlap in the emission, whereas 31\% are between the slow cooling synchrotron and fast cooling synchrotron lines, showing dominant non-thermal emission origins \citep{Acuner20, li2021, Wang2024}. 


Following \citet{li2021}, we calculated the correlation between $\alpha_{\max}$ and the pulse number. 
In our analysis of 61 pulses, we also see the same evolution with a mild correlation parameter $r_s$ of $-0.41$, further strengthening our hypothesis regarding the link between pulse shape and pulse emission evolution in a burst. We also looked at the correlation between the peak energy $E_{pk}$ and pulse shape as seen in Figure \ref{fig: epk_shape}. The $r_s$ parameter is 0.23 for the entire population, indicating a weak positive correlation.
These correlations suggest that as the relativistic shell evolves, the emission processes may change and could affect the shape of the light curves. 

\subsection{Pulse duration versus pulse shape relations}
In this section, we look at the effect of different pulse durations on the pulse shape and, consequently, on the underlying emission mechanisms. The plot contains 61 pulses from 22 GRBs mentioned at the beginning of section \ref{sec:obsrvations}. 

\begin{figure}
    \includegraphics[width=0.499\textwidth]{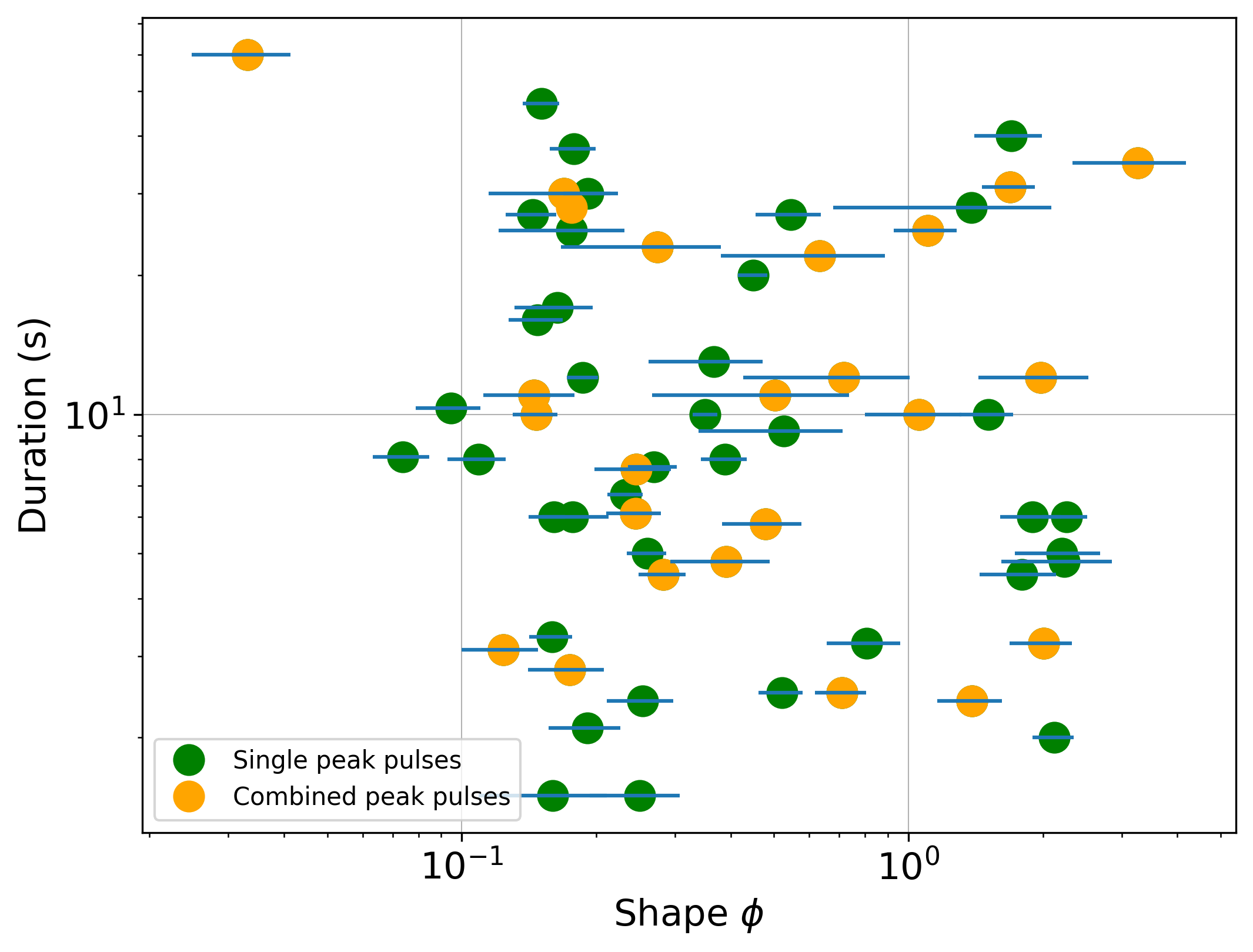}
    \hfill
    \includegraphics[width=0.499\textwidth]{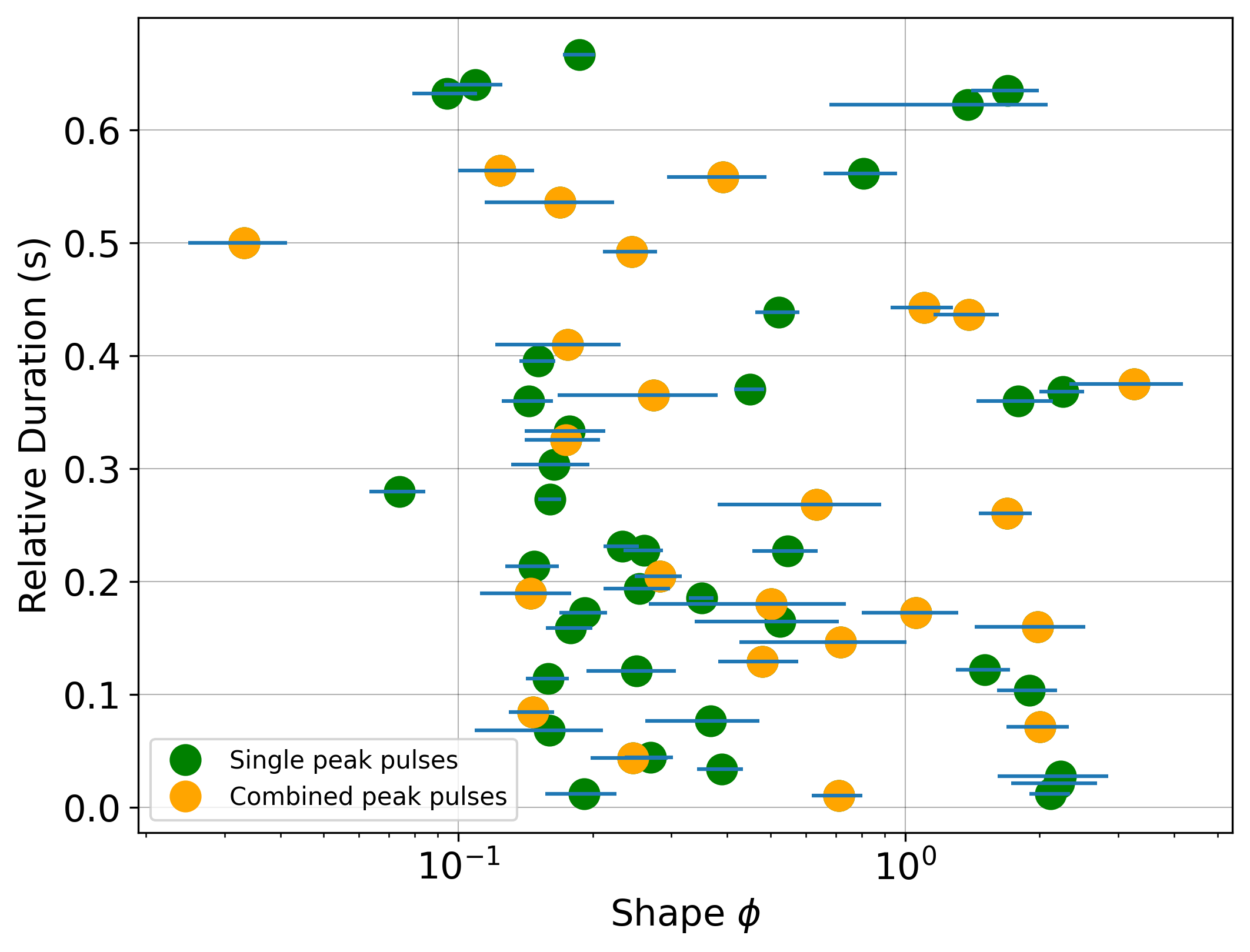}
\caption{Left: Pulse duration versus shape relation. Right: Relative pulse duration (${\rm pulse \ duration}/T_{90} $) versus shape relation. Green and orange dots represent single and combined peak pulses. Data points are the same as in Figure \ref{fig:shape_plsno_err}.} \label{fig: shape_duration}     
\end{figure}

In Figure \ref{fig: shape_duration} (left), the x-axis represents the pulse shape parameter $\phi$ after fitting with Equation \ref{eq:peer_ag}. The y-axis represents each pulse's duration in seconds. No noticeable correlation is found. 
For the correlation analysis for the pulse duration - pulse shape relation, Spearman's $r_s$ is $-0.11$ with a $\rm$ p-value $= 0.41$. There is no significant difference in Spearman's $r_s$ values for the single and combined peak pulse populations. The $r_s$ values are -0.19 with a p-value of 0.08 and 0.02 with a p-value of 0.92, respectively. This indicates that the duration of single peaked pulses and combined peak pulses are not correlated with the shape. 

Similarly, for the relation between relative duration - pulse shape in Figure \ref{fig: shape_duration} (right), the x-axis
represents the pulse shape parameter values after fitting with the novel pulse shape function. The y-axis represents the relative pulse duration in seconds of each pulse. The relative duration is defined as ${\rm pulse \ duration}/T_{90} $. Some noticeable trend was found here. The Spearman's $r_s$ value is $-0.22$ with a p-value = $0.09$. The Spearman's $r_s$ values indicate that the pulse duration and the time it takes to emit a single pulse w.r.t to total emission time have a mild effect on the shape of the light curve. A slight difference exists in the $r_s$ values in this correlation between single peak and combined peak pulses. The values are $-0.23$ and $-0.32$, respectively. This indicates that the relative duration of combined peak pulses is mildly correlated with the shape. \\

\begin{figure}[ht!]
    \centering
    \includegraphics[width=12cm]{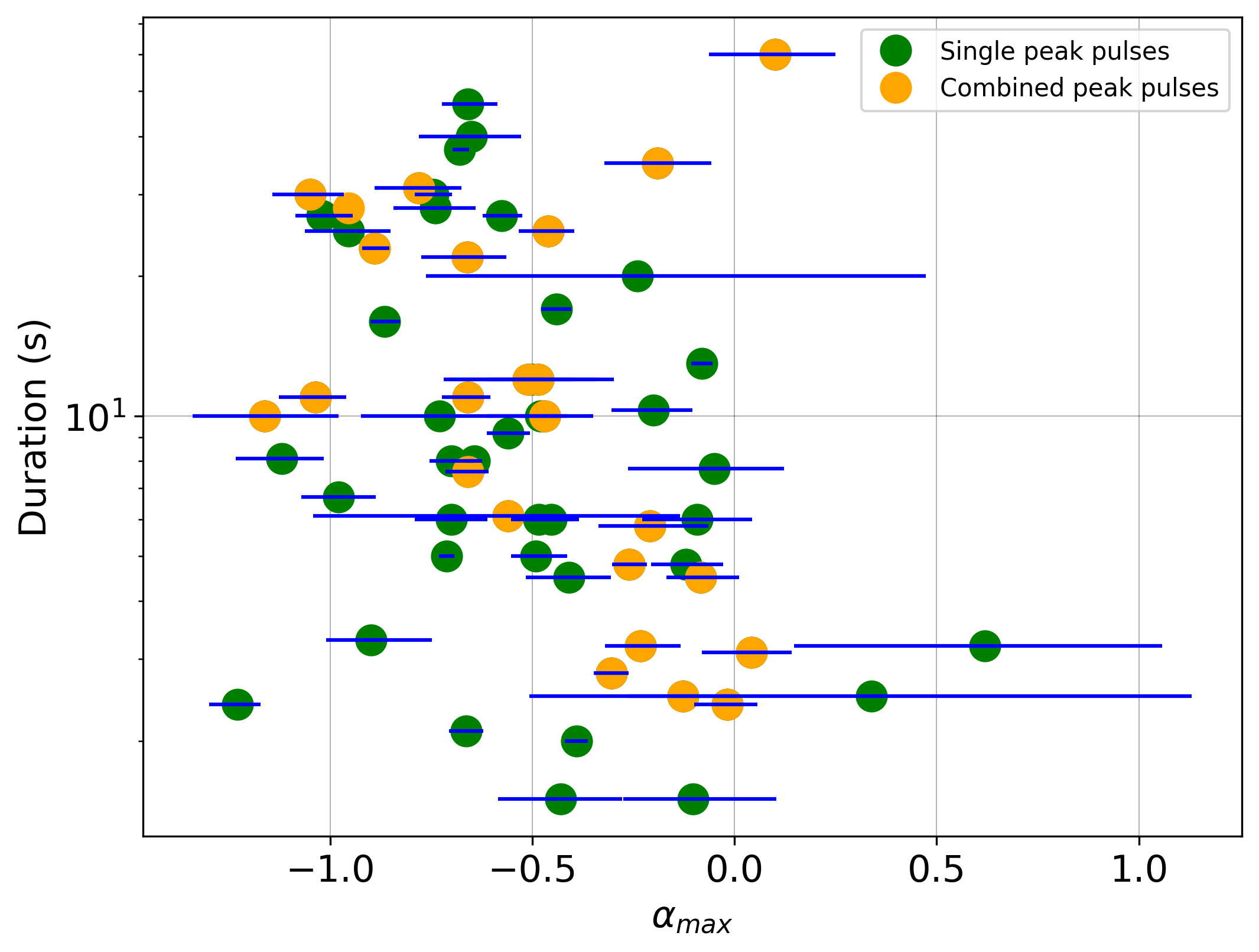}
    \caption{ Pulse duration versus $\alpha_{\max}$ relation. The green and orange data points are single peak pulses and combined peak pulses, respectively. Data points are the same as in Figure \ref{fig:shape_plsno_err}.}
    \label{fig:pulse_dur_alphamax}
\end{figure}

In Figure \ref{fig:pulse_dur_alphamax}, we plot all 61 pulses, and the x-axis represents the $\alpha_{\max}$ values of each pulse after the spectral fitting using the Band model.
The y-axis represents the pulse duration in seconds.
While the duration does not correlate with the pulse shape, surprisingly, here we see that for all the pulses, the Spearman's $r_s$  between $\alpha_{\max}- $pulse \ duration is $-0.36$ with a $\rm$ p-value $= 0.003$. For the single peak pulses, Spearman's $r_s$ value falls to $-0.29$ with a p-value = 0.08, and for combined peak pulses, the $r_s$ parameter is a stronger $-0.43$ with p-value = 0.04. This proposes that the duration has some correlation with the emission mechanism. 
The error bars for the pulse duration correspond to the detector's resolution, which is minimal, around 2 $\mu$s for TTE data during the burst. These correlations need to be further explored for better understanding.

\section{Summary and Discussion}

Our investigation aimed to integrate the time series and spectral studies to understand the inherent patterns present among the GRB light curves and its relation with underlying radiative mechanisms. 
We have chosen a sample set of 22 GRBs consisting of 61 pulses and proposed a novel function to quantify the
pulse shape characteristics. This was followed by a time-resolved `Band' model spectral analysis. Various correlations between pulse shape and spectral shape properties shed light on the radiation mechanism of the particles and the location of energy dissipation and prompt radiation in GRB jets.

The key results are as follows and further discussed below:
\begin{itemize}
\item We find a wide range of pulse shapes, from pulses that are symmetric to FRED shapes. 
We find that purely symmetric-like pulses are found in 26\% of all pulses, i.e., about 1/4 of the total pulses in our sample, which is much more ubiquitous than previously assumed. FRED shape pulses are found in 51\% (about 1/2) of the pulses in our sample, and the rest show a mixed shape.

\item The pulse shape $\phi$ (Equation \ref{eq: shape}) value decreases from symmetric-like (larger value) to FRED-like (smaller value) as the pulse number increases (see Figure \ref{fig:plsavg}). We find a mild correlation with Spearman correlation coefficient  $r_s = -0.34 $, as seen in Figure \ref{fig:shape_plsno_err}.
    
\item The spectral shape indicates a relation with the pulse shape as given in Figure \ref{fig:alpha_shape_emi}: the value of the low energy spectral index, $\alpha_{\max}$, of the pulses shows a trend with the pulse shape $\phi$ with Spearman correlation coefficient $r_s = 0.23$.
Pulses tend to become thermal-like (larger $\alpha_{\max}$), as their shape becomes more symmetric-like (larger $\phi$). 

\end{itemize}


\subsection{Pulse symmetry} \label{sub:pulse symmetry}
The pulse analysis method introduced in subsection \ref{sec:pulsemodel}  has been successful in quantifying the temporal symmetry observed in light curves. We are unaware of any instrumental bias
capable of causing small amplitude deviations in a FRED curve that would mimic the bright, long features seen as a symmetric-like curve. 
From Figure \ref{fig:shape_plsno_err}, one sees that many of the pulses in our sample are symmetric-like and that their fraction is particularly high for the initial pulses in a given burst.


The finding that a large fraction of the observed pulses in the prompt GRB light curve are symmetric is surprising.
The prompt emission is produced as a result of energy dissipation processes that occur in relativistic jets. 
The rise time scales of the pulse
indicate a continuous dissipation of energy: as more and more energy is dissipated, more
and more photons are emitted. On the other hand, the decay is subject to various light aberration effects, such as high-latitude emission. There is, therefore, no apriori reason to obtain a symmetric pulse light curve, as is observed. 

Several authors \citep{hakkila2018,hakkila2019,hakkila2021,liu2023} applied a fitting approach that combines a monotonic component with fast-varying residual structures. They found that these residuals exhibit temporal symmetry, with patterns in the decay phase appearing as time-reversed counterparts of those in the rise phase, typically reflected about the pulse peak time. In our sample of GRB lightcurves, we also identify time-reversed structures in the rising and decay phases of the pulses. We account for these underlying variations in the pulse definition itself and by the inclusion of the combined peak pulse category. 

To explain the symmetric nature of these structures, \citet{hakkila2019} suggested that some aspects of the pulsed emission may be repeated in reverse order as a result of events in the jet; either the emission mechanism must take this into account through the jet’s kinematic behaviour or a different emission mechanism is required to explain the
time-reversed portion of the light curve. 

An additional clue may come from spectral analysis. 
Considering the results from Figure \ref{fig:alpha_shape_emi}, one finds that the $\alpha_{\max}$ values tend to decrease as pulse number increases and time evolves. As discussed above, large values of  $\alpha_{\max}$ are indicative of a possible contribution from a thermal (photospheric) component. The decrease of the value of $\alpha_{\max}$ with the pulse shape may, therefore, indicate a change in the emission mechanism as the emission evolves. This raises the possibility that the similar timescales of rise and decay in a symmetric pulse are due to diffusion effects below the photosphere. Further supporting this possibility, we note that \citet{Acuner2019} concluded that around a quarter of all bursts have pulses that are from non-dissipative outflows, purely based on spectral studies. This is a similar fraction of pulses that we find to be symmetric. We further discuss the theoretical consequences of this possibility below. 

The majority of the pulses in our sample lie between the slow-cooling synchrotron and the NDP line, above which only a photospheric component is theoretically expected. Intermediate values can, therefore, be explained as a simultaneous contribution from both thermal (photospheric) and non-thermal (synchrotron) emission. 
Harder $\alpha_{\max}$ values of the initial symmetric-like pulses in a given burst can be explained as a stronger contribution of thermal emission components in the overall spectra. 
The more FRED-like pulses have soft $\alpha_{\max}$ values, indicating non-thermal emission origin. This result is a natural outcome of the classical GRB ``fireball" model: thermal emission originates from the photosphere, which is the closest radius to the progenitor from which photons escape and is thus expected to be the first signal detected. 
 Non-thermal spectra are possibly emitted from larger radii (well above the photosphere), resulting in more FRED-like light curves. We conclude that a different emission radius is the most likely origin of the various pulse shapes.


\subsection{Pulse duration effects on the shape}
From the results presented in Figure \ref{fig: shape_duration}, we conclude that the pulse duration does not correlate with the pulse shape and, by extension, the pulse emission
mechanism. This seems counterintuitive since the pulse durations in our sample vary by one order of magnitude (from a few seconds to $>50$ s). This indicates a similarity between different pulses, hinting towards a common underlying origin. One caveat is that the redshifts of most GRBs in our sample (all except four) are unknown. Therefore, we could not correct for the different redshifts.

Due to the binning of pulse count-rate with a limiting error in the order of 0.1~s, the novel pulse shape model is not sensitive to small, fast variations in the pulse light curves. 
The typical pulse duration of a few seconds represents a characteristic time scale within the progenitor, e.g., fluctuation within the accretion. On the other hand, a shorter time scale ($\sim$ ms) may represent local variations in the flow, e.g., due to turbulent motion. Our choice of integrating between hundred to two hundred time bins for the pulses of these long GRBs should, therefore, be sufficient to capture the global properties of the outflow. 

Differences are found in the Spearman correlation coefficients for single-peak pulses and combined-peak pulses when considering the relative duration-pulse shape relation presented in Figure \ref{fig: shape_duration}. This could be explained by the structural difference in combined peak pulses.
Combined peak pulses are constructed of either a single pulse with random fluctuations of relatively small
magnitude (whose physical origin is uncertain) or made of separate pulses with very similar shapes. Even if this is the case,
these pulses are close to each other, likely emerging from the same region.

Initially, results from prompt emission pulses observed by BATSE showed that the pulse width remains constant throughout the GRB time history \citep{ramirez2000}. However, using our Fermi-GBM dataset, we now show that later pulses tend to be longer in duration than earlier ones, with a Spearman correlation coefficient $r_s$ = 0.25 observed between the pulse duration and pulse number. These findings are consistent with the results reported by \cite{hakkila2024}. This, along with the hard-to-soft evolution, appears consistent with the known properties of X-ray flares \citep{chincarini2007,chincarini2010,margutti2011,dereli2025}.

\subsection{On the novelty of the fitting function}
 
As a fitting function, we consider two approximate sigmoid functions that independently fit for the rising slope $s_l$ and decaying slope $s_r$. The half-time of the rise phase (pulse start to peak) and the decay phase (peak to pulse end) is represented by $r_l$ and $r_r$, respectively (see  Equation \ref{eq:peer_ag}). In our function, these are treated as independent parameters. This is the key difference from previous fits \citep{Norris1996, hintze2022,norris2005}, which enables the provision of good fits to wide pulses whose peak extends over a substantial duration. This additional degree of freedom is crucial in obtaining symmetric pulse shapes.
For optimal fits, we always require $r_l \le r_r$; otherwise, the obtained slope values do not provide insight into the observed ones. This condition imposes a restriction on the fits, which, however, do not affect the overall conclusion. 

Due to our pulse definition criteria, we managed to extract a large number of pulses from these bursts compared to previous studies. The pulse definition we use is purely dependent on the S/N ratio of the count rates. This definition may differ from the criteria that \citet{hakkila2014,hakkila2018} and \citet{hakkila2021} applied to the BATSE pulse identification, which further considers the residual structures along with the monotonic pulse fit.  It is also possible that the $ 2\mu s$ resolution for the GBM TTE data might be influencing the pulse identification because of the increased sensitivity to variations, causing larger pulse numbers. These factors might be the cause of the difference in the number of pulses found in the two analyses.\\

While previous studies have made the delineation between pulse and structures \citep{hakkila2021, liu2023}, there is not yet a consensus on a physical model of the emission mechanism requiring these structures. Hence, here we pursue an alternative, empirical perspective of pulses, which is mathematically well-defined and reproducible without imposing any assumption of pulse morphologies on the data.
Due to the flexible approach of the pulse definition used in this work, we are able to capture a broader range of pulse shapes for both single and multipulse GRBs, which can include variations in the pulse shapes. This offers a complementary approach to previous light curve studies.

We acknowledge that the flexibilities in our pulse definition allow for a broader range of pulse morphologies. In future studies, it will be interesting to compare how the correlation results obtained from different pulse definitions depend on the initial selection criteria. Although we emphasize that the trends we see in this study are self-consistent within the framework we use.

\subsection{Theoretical interpretation}
The symmetric nature of pulse shapes warrants an investigation into the jet dynamics.
By combining the results presented in subsections \ref{sub: pulse shape studies} and \ref{sub: spectral shape studies}, we infer that the initial symmetric-like pulses have stronger thermal emission components originating from the photosphere.
Above the photosphere, pulses are expected to be asymmetric as the rise and decay phases are governed by different physical mechanisms. Below the photosphere, photon diffusion can result in symmetric pulses. However, the challenge lies in the timescale: the typical timescale for radiation from the photosphere to be detected is $r_{ph}/(\Gamma^2 c)$. The photospheric radius  $r_{ph}$ is given by \citep[e.g.,][]{peer2012},
\begin{equation}
r_{ph} = {L \sigma_T \over 8 \pi m_p \Gamma^3 c^3} = 5.8 \times 10^{12}~L_{52} \Gamma_2^{-3}~{\rm cm}        
\end{equation}
Here, $L$ is the GRB luminosity, $\Gamma$ is the jet Lorentz factor, $\sigma_T$ is the Thomson cross section, $m_p$ is the proton mass, and $c$ is the speed of light. 
For typical values of outflow parameters, $\Gamma \gtrsim 10^2$, the photospheric radius is in the range $ 10^{11} - 10^{13}$~ cm, and the resulting timescale is only a fraction of a second. This timescale is shorter by 2-4 orders of magnitude than the typical pulse duration in our sample, which is about a second or more. However, this time scale is extremely sensitive to the uncertain value of the jet Lorentz factor, as it is proportional to $\Gamma^{-5}$. Therefore, if the typical Lorentz factor of GRBs in our sample is lower than the commonly assumed value and is rather of the order of only a few tens, as recently suggested \citep{dereli2022}, this discrepancy disappears. 
The results presented here may, therefore, provide additional, indirect evidence that the Lorentz factor of many GRBs may be only a few tens rather than a few hundreds, as is commonly assumed. For further indications on low values of GRB Lorentz factors, see \cite{dereli2025}. 

From the discussion in section \ref{sub:pulse symmetry}, we also infer that the later FRED-like pulses that show non-thermal emission components originate further away from the progenitor.
According to synchrotron theory, the spectral peak energy, $E_{pk}$ is a function of the jet Lorentz factor, $\Gamma$, the (comoving) electrons characteristic Lorentz factor, $\gamma_m$, and the magnetic field, $B$. Following the initial acceleration phase (whose observed duration is a fraction of a second) the values of all these parameters are either constant or decreasing with radius. Therefore, one naturally expects lower values of $E_{pk}$ of pulses originating at larger radii.
As we discussed above, emission originating at larger radii, is expected to have a more FRED-like pulse shape. Indeed, when looking at Figure \ref{fig: epk_shape}, one finds a trend suggesting lower $E_{pk}$ values for pulses that have a more FRED-like shape. 

The overall emerging picture, therefore, indicates a range of radii where energy dissipation occurs.
Early pulses are close to the photosphere, while later pulses occur further above it. 
This implies either an evolution in the central engine activity over time or a similarity in material ejection by the central engine. 
Otherwise, one would expect the pulse shapes to be randomly distributed rather than showing a systematic progression from symmetric to FRED-like pulses. The results obtained, therefore, strongly suggest both (i) that the GRB pulses are produced at a range of radii, some below and some above the photosphere, and (ii) that the GRB jet Lorentz factor in many GRBs is $\Gamma < 100$, possibly in the range of a few tens.

\section*{Acknowledgments}
GA thanks Dhruva Sambrani and Damien B\'egu\'e for the helpful discussion and comments. We also thank the anonymous referee for suggestions that improved the clarity of the manuscript.
This research made use of the High Energy Astrophysics Science Archive Research Center Online Service HEASARC at the NASA/Goddard Space Flight Center.
AP acknowledges the support from the European Union (EU) via ERC consolidator grant 773062 (O.M.J.), and from the Israel Space Agency via grant \#6766. FR  acknowledges
support from the Swedish National Space Agency (2021-00180 and 2022-00205). 
\newpage

\bibliography{sample631}{}
\bibliographystyle{aasjournal}
\newpage
\renewcommand\thefigure{\thesection.\arabic{figure}}
\setcounter{figure}{0}
\appendix

\section{Comparison of the fitting function used here and in Norris et al. (2005)} 
\label{sec:Appendix}
Here we compare the fits carried out using Equation \ref{eq:peer_ag} with fitting functions that appear previously in the literature, specifically the one used in \cite{norris2005}.   
We applied the \cite{norris2005} model to our sample of 61 pulses from 22 GRBs. 
To ensure consistency, we have applied the same sample selection processes to all GRB pulses. 
To demonstrate the advantage of Equation (\ref{eq:peer_ag}), we give here two examples of pulse shape fits.

\begin{figure}[h]
\centering
  \centering
  \includegraphics[width=.45\linewidth]{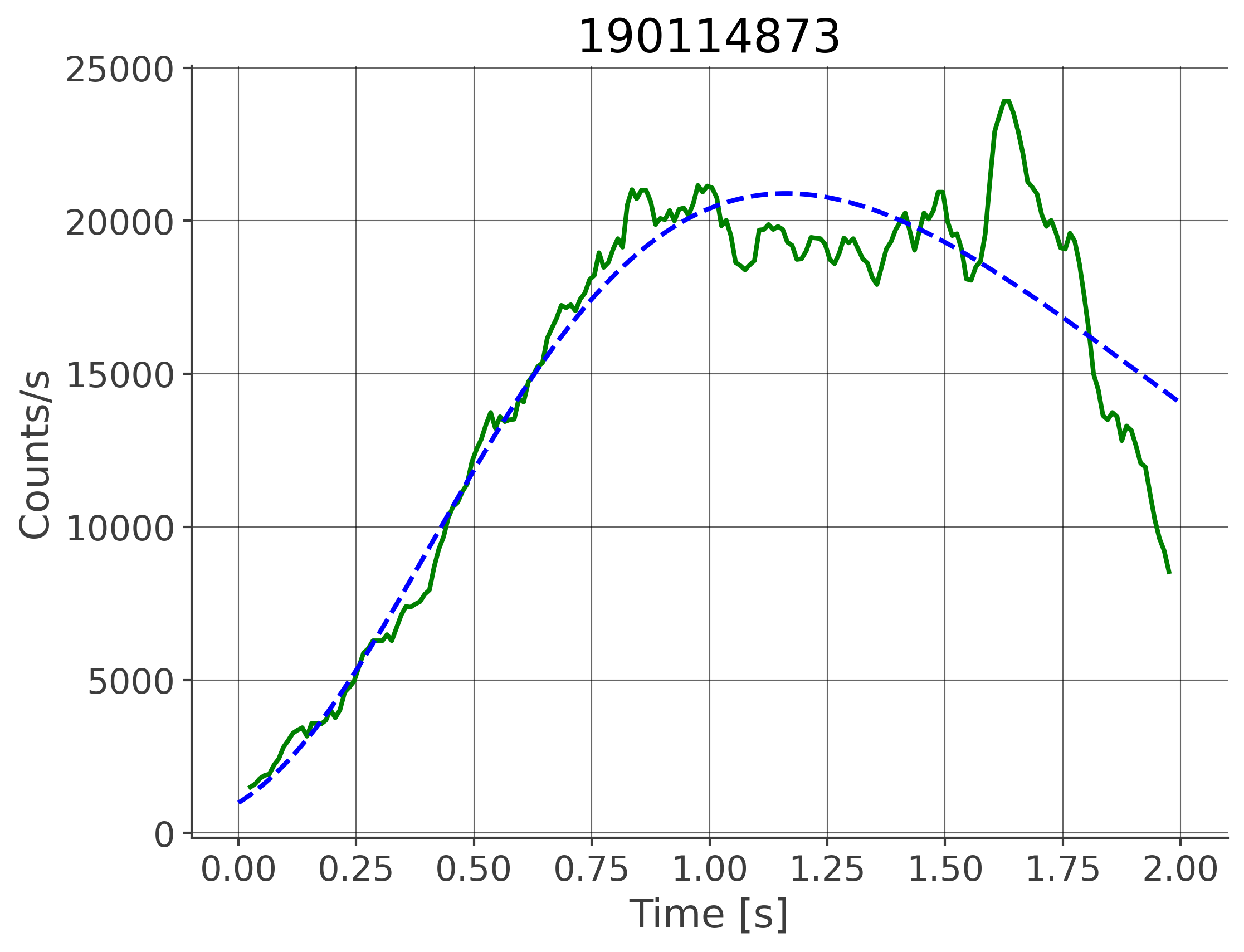}
   \includegraphics[width=.45\linewidth]{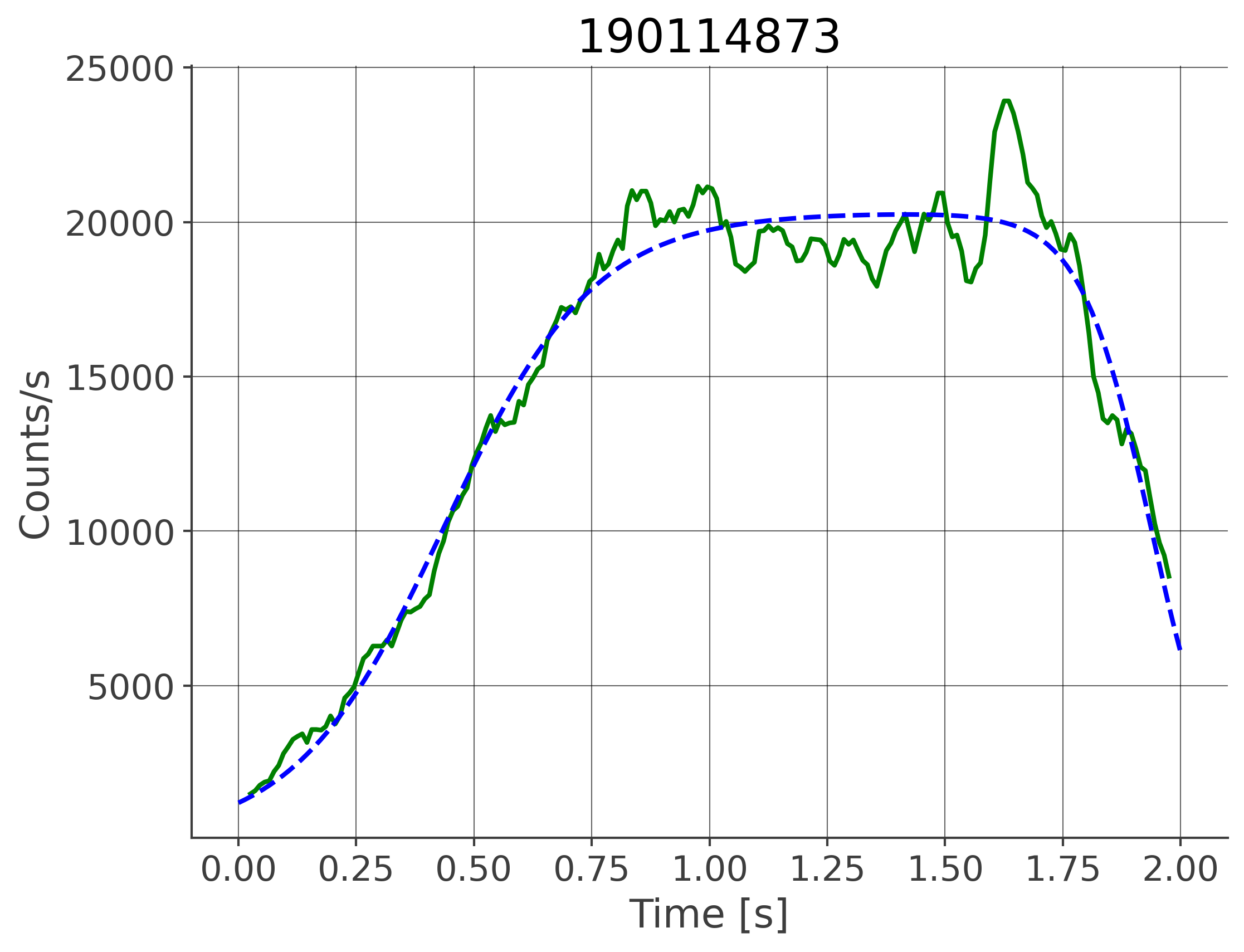}
  \caption{Comparing fit of GRB190114 using the Norris et. al. (2005) function [left] and the function used here [right]. The solid green line represents the raw count-rate data from GBM, and the dashed blue line represents the pulse shape model fit according to \cite{norris2005} Function and Equation \ref{eq:peer_ag}, respectively.}
\label{Fig:1}
\end{figure}
\begin{figure}
\centering
  \centering
  \includegraphics[width=.45\linewidth]{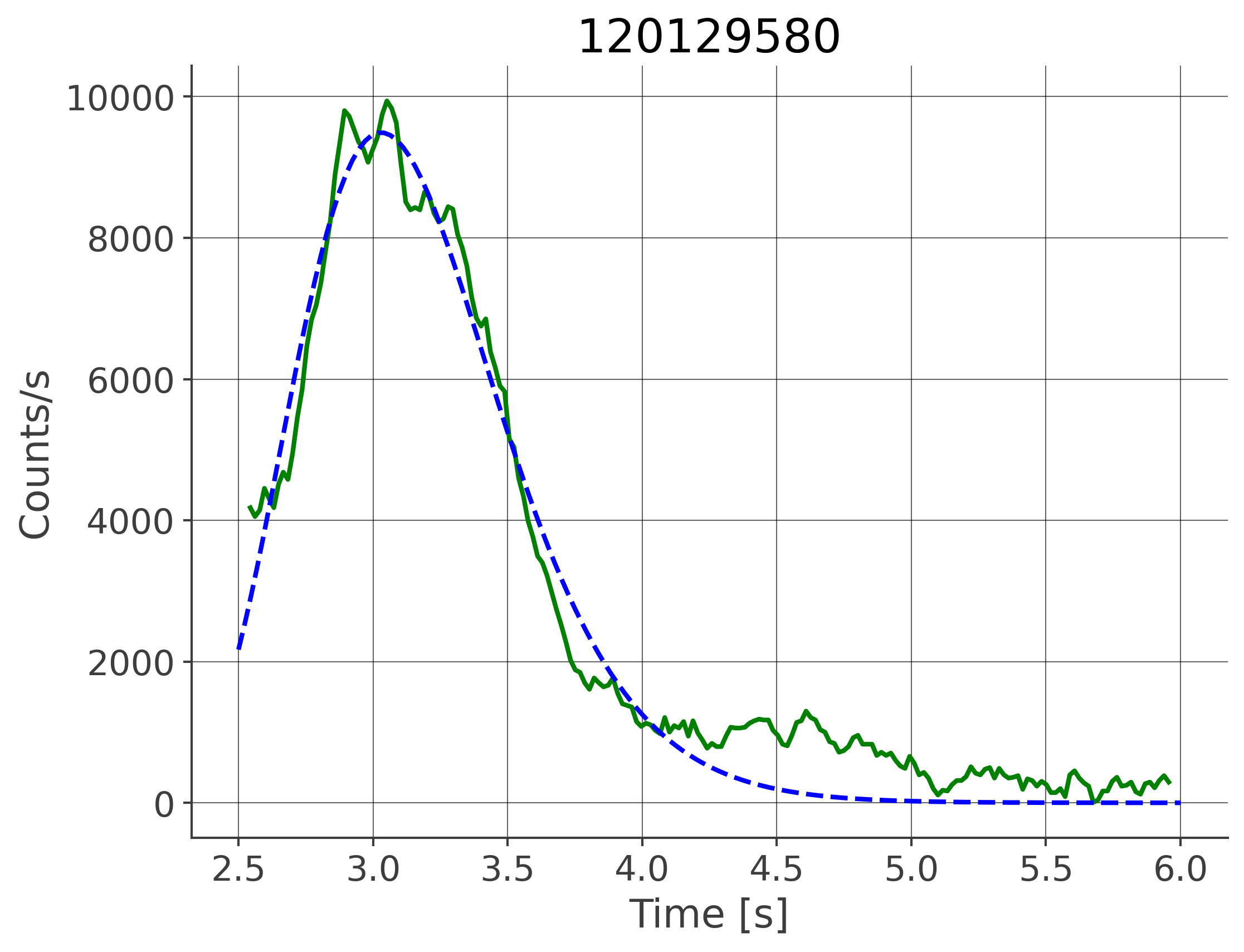}
   \includegraphics[width=.45\linewidth]{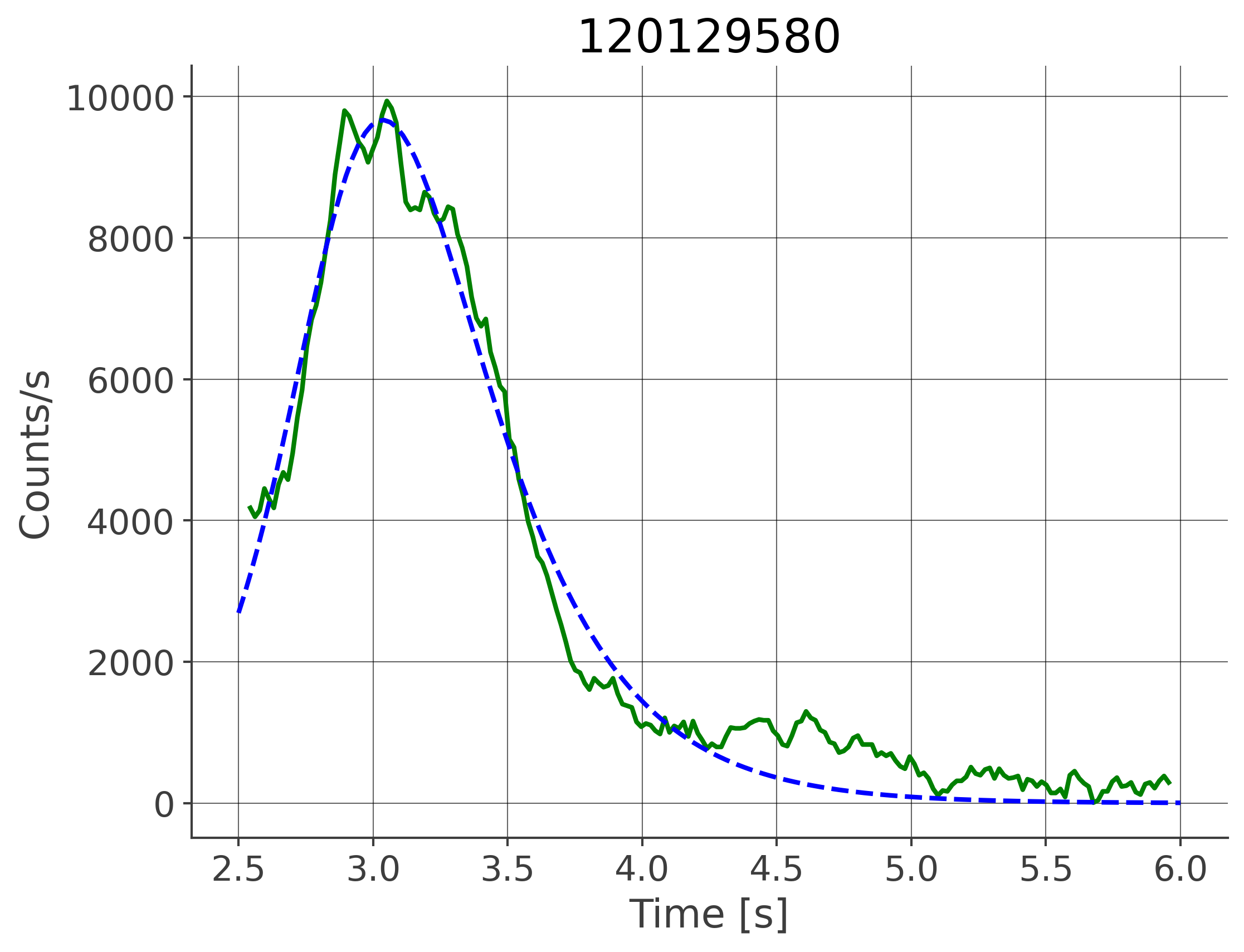}
  \caption{Comparing fit of GRB 120129 using the Norris et. al. (2005) function [left] and our function used here [right]. The solid green line represents the raw count-rate data from GBM, and the dashed blue line represents the pulse shape model fit according to \cite{norris2005} Function and Equation \ref{eq:peer_ag}, respectively.}
\label{Fig:2}
\end{figure}

In Figure \ref{Fig:1} we present the example of GRB 190114 fitted twice: once with the function used by \cite{norris2005} [left] and using Equation (\ref{eq:peer_ag}) [right]. This lightcurve shows a steep decay, which is better captured using our Equation (\ref{eq:peer_ag}): The \cite{norris2005} function fit gives $k=0.53 \pm 0.02$ while we obtain $\phi = 2.12 \pm 0.22$.
The $r^2$ values are similar - 0.89 and 0.94 for the two fits, respectively. 
As is demonstrated in Figure \ref{Fig:1}, Equation (\ref{eq:peer_ag}) is more flexible in fitting pulses with longer rise times compared to decay times.

A second example is given by fitting the pulse of GRB 120129, shown in Figure \ref{Fig:2}. While both the \cite{norris2005} function [left] and Equation (\ref{eq:peer_ag}) [right] provide good fits to the pulse shape - both give $r^2 = 0.96$, we point out that the Norris function fit gives $k = 0.57 \pm 0.03$, which is the same value as obtained for fitting the pulse of GRB 190114 shown in Figure \ref{Fig:1} above; while in the fit carried using our Equation (\ref{eq:peer_ag}), we obtain $\phi = 0.51 \pm  0.06$, which is different than the value obtained above, as is evident from the pulse shape. Due to the wider range of possible $\phi$ values compared to the $ k$  values, the fits of Equation (\ref{eq:peer_ag}) can provide more distinctive values for each pulse shape. 

Furthermore, applying the \cite{norris2005} function to our sample set, we found that 12 pulses did not meet the goodness-of-fit criteria $r^2 > 0.7$.
We further found that 7 pulse shape fits were slightly better with the \cite{norris2005} model compared with Equation (\ref{eq:peer_ag}). 
However, the maximum improvement in goodness-of-fit $r^2$ was only $\sim 2.75\%$. For example, the \cite{norris2005} function fit of GRB 131014 provides a fit with $r^2 = 0.92$, whereas the function we use gives  $r^2 = 0.90$. In all other pulses, Equation (\ref{eq:peer_ag}) produced better fits.

\end{document}